\documentclass[aps,prfluids,amsmath,amssymb]{revtex4-2}

\usepackage{graphicx}
\usepackage{dcolumn}
\usepackage{bm}
\usepackage{amsmath}
\usepackage{xcolor}

\newcommand{\red}[1]{\textcolor{black}{#1}}     
\newcommand{\blue}[1]{\textcolor{black}{#1}}    
\newcommand{\orange}[1]{\textcolor{black}{#1}}  

\begin{document}

\preprint{APS/123-QED}

\title{Scale-resolving simulations and data-driven modal analysis of turbulent transonic buffet cells on infinite swept wings}

\author{David J. Lusher}
 \email{lusher.david@jaxa.jp}
  \author{Andrea Sansica}%
\affiliation{Japan Aerospace Exploration Agency (JAXA), Aviation Technology Directorate, Chofu Aerospace Center, 7-44-1 Jindaiji Higashi-machi, Chofu-shi, Tokyo 182-8522, Japan}%

\date{\today}

\begin{abstract}
\blue{Transonic buffet} is a class of shock-wave/boundary-layer interaction (SBLI) known to exhibit self-sustained two-dimensional (2D) chordwise shock wave oscillations (Strouhal number $St \sim 0.05-0.1$), and three-dimensional (3D) spanwise-modulated flow separation/reattachment ($St \sim 0.2-0.4$). Due to computational cost, scale-resolving simulations of span-periodic configurations to date have been limited to narrow airfoils (aspect ratio $AR = L_z / c \sim 0.05-0.25$, for span-width $L_z$ and chord $c$). These ratios are insufficient to accommodate the 3D `buffet cell’ instability reported in low-fidelity simulations and experiments. 
In this work, implicit large-eddy simulations (ILES) and modal analysis are performed on infinite wings up to $AR=3$ with \blue{sweep angles between $\lambda = \left[0^{\circ}, 35^{\circ}\right]$}. The sensitivity of the 2D and 3D modes to crossflow is detailed. Two flow conditions are examined, corresponding to minimally and largely separated mean flow at the shock location. For the minimally separated case, the shock dynamics remain essentially spanwise-uniform (quasi-2D), with only weak and intermittent separation cells confined to the trailing-edge region and exhibiting negligible interaction with the shock. In contrast, increased mean separation leads to the emergence of pronounced 3D buffet cells with a characteristic spanwise wavelength $\lambda_z \approx 1$–$1.5c$. Spectral proper orthogonal decomposition reveals that a quasi-stationary low-frequency 3D separation mode previously identified on unswept wings ($St \approx 0.02$) becomes a spanwise travelling mode as sweep is imposed, shifting monotonically to intermediate frequencies ($St=0.06$–$0.35$). The 2D shock mode is largely insensitive to sweep, whereas the frequency and energy content of the 3D mode increase with sweep while its wavelength remains unchanged. The results demonstrate that transonic buffet \blue{on infinite wings} arises from the superposition of distinct but \red{coexisting} 2D shock motion and separation-driven 3D instabilities, with mean flow separation at the shock identified as a necessary condition for dominant 3D buffet dynamics to emerge.
\end{abstract}

\maketitle


\section{Introduction \label{sec:introduction}}

Turbulent transonic buffet is an important type of self-sustained shock wave/boundary-layer interaction (SBLI) known to occur under specific flight conditions characterised by high subsonic Mach numbers and moderate-to-high angles of attack \citep{L2001}. Buffet results in strong, self-excited oscillations of lift and drag, which can have severe detrimental effects on the structural integrity and operational flight envelopes of aircaft, and therefore, must be mitigated and controlled in a wide range of aerospace applications. The unsteady transonic flow field around the wing consists of regions of both subsonic and supersonic flow. Type II buffet on supercritical wings \citep{Giannelis_buffet_review} has been shown to consist of two instabilities: (i) two-dimensional (2D), low-frequency, chord-wise oscillations of the main shock wave at Strouhal numbers of $St \sim \left[0.05, 0.09\right]$, and (ii) three-dimensional (3D), spanwise cellular separation bubbles known as buffet cells \citep{IR2015}, which typically occur at Strouhal numbers approximately one order of magnitude higher ($St \sim \left[0.25, 0.45\right]$) than the 2D mode \citep{PDL2020} in swept wing configurations. Buffet cells are typically identified as outboard propagating pressure fluctuations along the spanwise length of the shock \citep{OIH2018}. 

While the relative importance of shock waves and flow separation in buffet is still one of the fundamental open questions, the relation between these two instabilities, their relevance for realistic 3D configurations and the mechanisms regulating them remain unclear. Investigating the flow and geometrical conditions under which these instabilities exist (and co-exist) is of the utmost importance to understand this complex phenomenon.

To isolate the 2D instability, numerical simulations on 2D airfoils and 3D span-periodic (infinite) wings with a narrow spanwise extent (typically less than $5-10\%$ of the chord length) have been extensively investigated. Given the relative lower computational cost, the literature is vastly populated by studies on these simplified configurations, which served a useful and detailed characterization of the 2D-chordwise shock oscillation unsteadiness. Reynolds-Averaged Navier–Stokes (RANS)-based studies constitute the largest body of literature in this context and have been employed to investigate turbulence model sensitivity \citep{GH2004,TC2006}, the origin of the driving mechanisms \citep{CGMT2009,IR2012,SMS2015,CGS2019,PL2019}, the effect of Mach number \citep{GLV2018}, regions of effectiveness for active control strategies \citep{PMSRD2019}, and to propose data-driven decompositions and reduced-order models \citep{PRD2019,SLKHR2022,KAWAI_RESOLVENT2023}. Hybrid RANS/large-eddy simulation (LES) approaches, such as detached eddy simulation and related variants \citep{D2005,GBH2014,IIHAT2016}, have also been successfully applied and shown to reproduce the dominant 2D buffet dynamics. More recently, advances in high-performance computing (HPC) have enabled a growing number of high-fidelity investigations that remove the error introduced by approximate turbulence models, instead applying LES and direct numerical simulations (DNS) to the problem of airfoil buffet \citep{GD2010,FK2018,zauner2019direct,Moise2023_AIAAJ,LongWong2024_Laminar_buffet,LSSMSH_OpenSBLIv3_CPC2025,lusher2025highfidelity_JFM}. Aside from a few exceptions \cite{LSSMSH_OpenSBLIv3_CPC2025,lusher2025highfidelity_JFM}, these buffet studies have typically been restricted to narrow spanwise domain widths ($5-10\%$ of airfoil chord length) and laminar airfoils \citep{dandois_mary_brion_2018}, for which buffet cells have not been observed \citep{moise_zauner_sandham_2022}. As the topic of the present study is the 3D transonic buffet cell mode established under turbulent conditions, we omit further discussion of both narrow and laminar airfoils, and low $\mathcal O\left(10^3-10^4\right)$ Reynolds number effects.

For realistic 3D geometrical configurations, experiments and RANS/URANS-based computations have been carried out on extruded unswept and swept wings mounted on side walls \citep{DSO2021,DSO2022,SKK2022,SHKK2022}, finite wings \citep{OIH2018,Houtman_Timme_Sharma_2023}, and full-aircraft configurations \citep{ST2017,hashimoto2018current,SKNNNA2018,MTP2020,T2020,SNKNNA2021,SH2023}. Consistently, these studies indicate that transonic buffet is an inherently 3D phenomenon, with the dominant mechanism associated with the cross-flow convection of buffet cells. When finite-wing effects are taken into account, the 2D shock oscillation instability weakens or disappears entirely, and transonic buffet becomes essentially a 3D phenomenon. The experiments of \citet{DSO2021,DSO2022} provide valuable insight into the coexistence of the 2D and 3D instabilities. When assessing extruded wings with both ends mounted on side-walls, they showed the presence of both instabilities. However, when finite-wing effects were considered, the authors showed that the 2D shock oscillation instability is significantly reduced, and the flow dynamics become dominated by the 3D buffet-cell instability. Despite general agreement on the absence of the 2D instability on finite-wing configurations, recent global stability analyses (GSA) \citep{SH2023} and pressure-sensitive paint (PSP)-based modal decompositions \citep{VMD_Ohmichi2024} suggest that a 3D counterpart of the 2D chord-wise shock oscillation may emerge at deep buffet conditions in the form of a quasi-2D, long-wavelength mode. 

\begin{figure}
\begin{center}
\includegraphics[width=0.497\columnwidth]{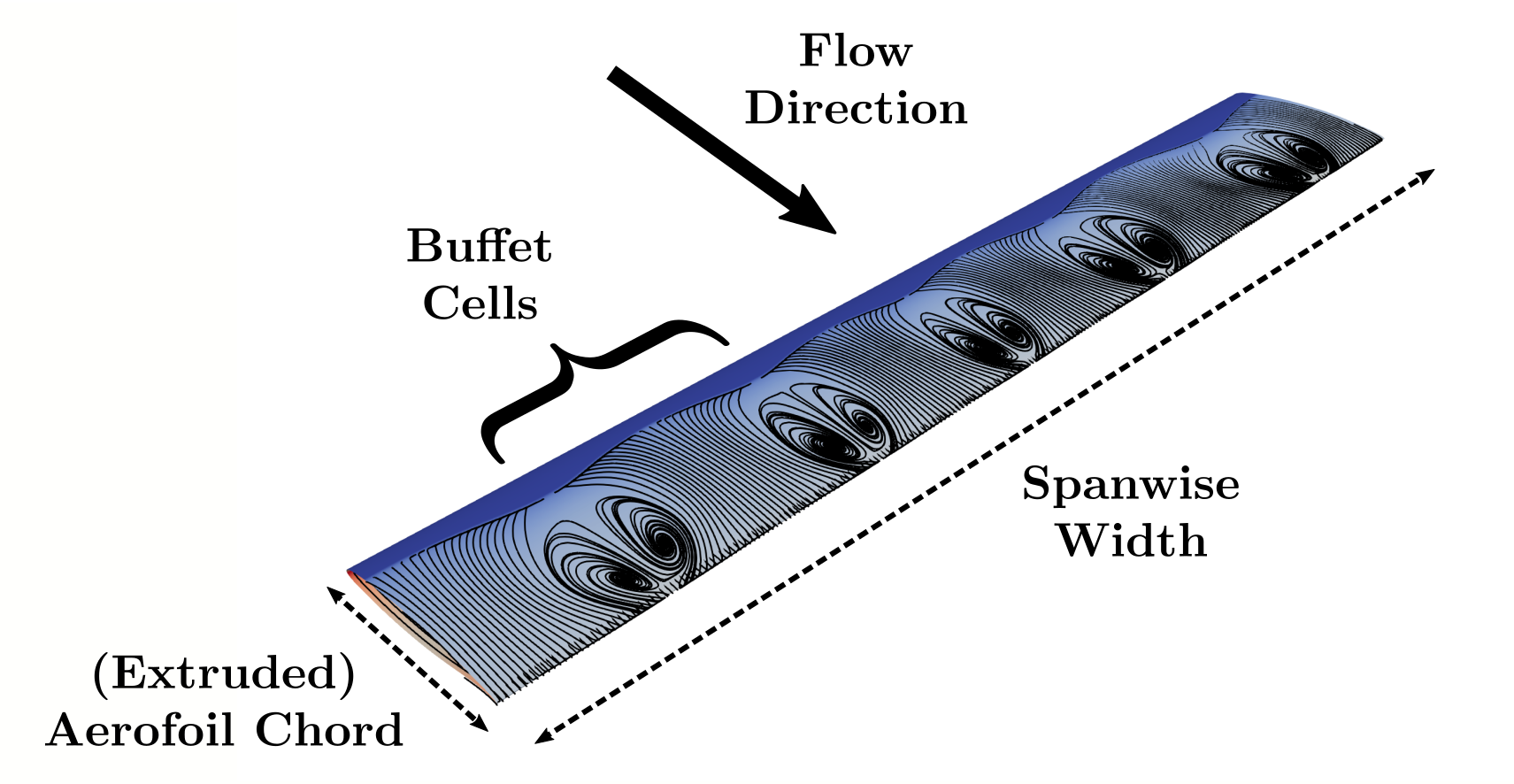}
\includegraphics[width=0.497\columnwidth]{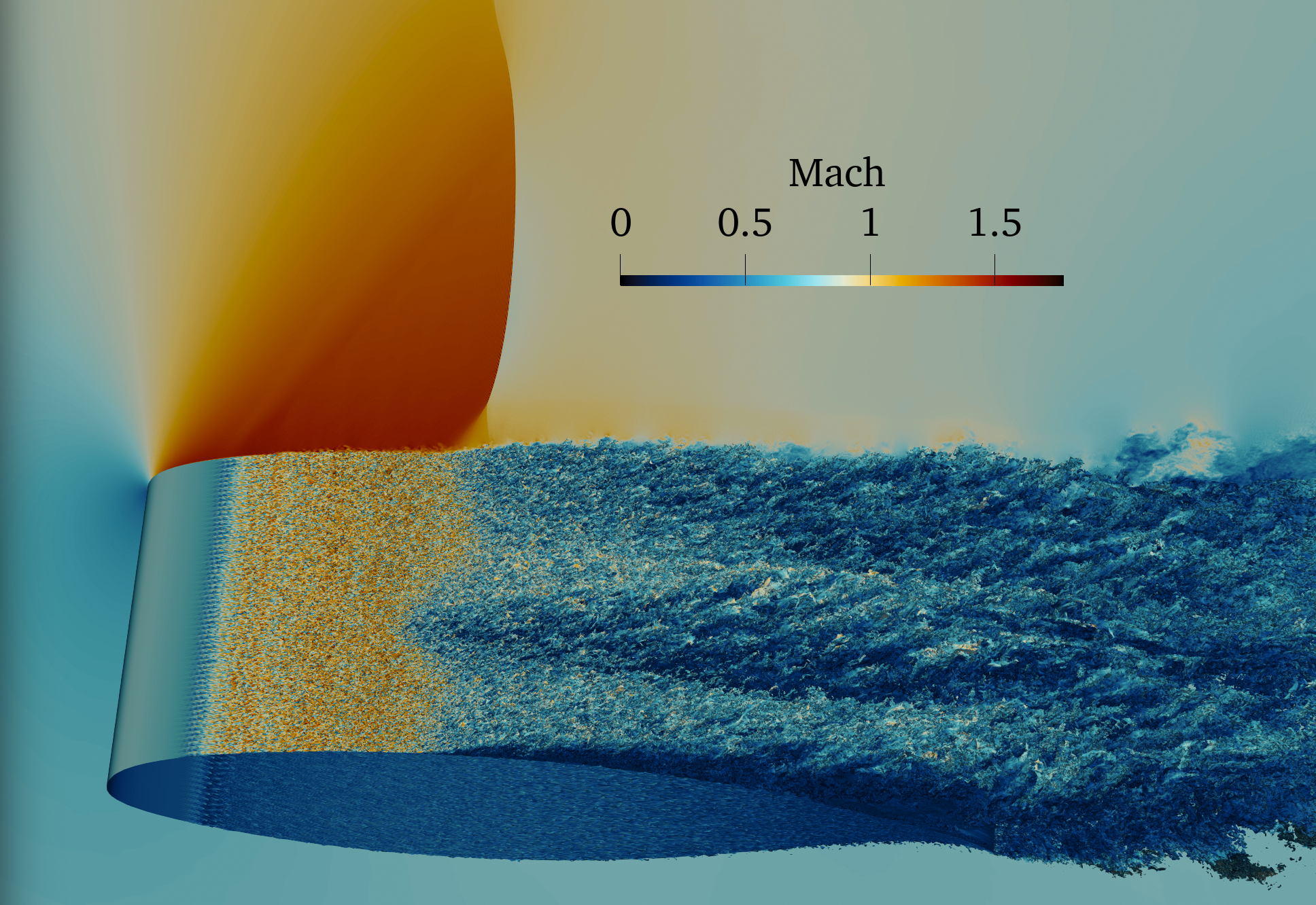}
\end{center}
\caption{(left) Example spanwise separation `buffet cells' in a low-fidelity steady RANS-based transonic airfoil simulation and (right) a high-fidelity simulation of the same phenomenon in the present study at a wing aspect ratio $AR = L_z / c$ equal to three chord lengths. Scale-resolved fine-scale turbulent structures at $w \pm 0.075$ are coloured by Mach number. The $\lambda$-shock wave is also visible on the Mach number contours imposed on the back panel. The flow is tripped to turbulence at 10\% chord.}
\label{fig:CRM_3D}
\end{figure}

Nevertheless, disentangling fundamental buffet mechanisms from purely geometrical effects remains a considerable challenge. Straight-wing experiments are inherently influenced by wind-tunnel side walls, which impose additional three-dimensionality on the problem through flow confinement and corner separations that interact with the primary shock system. 
It should be noted that while attempts to realise nominally-2D buffet experimentally have been made \citep{Jacquin2009,SKNNNA2018,Brion2020_Laminar_Experiment}, it remains difficult to quantify the impact of residual three-dimensionality, especially on wings with low aspect ratios. Manufacturing large aspect ratio wings is in fact challenging for assuring structural integrity during the test and accommodating the necessary instrumentation. In full-aircraft geometries, even in simplified configurations without high-lift devices or engine nacelles, wing–body junction separations are present, further complicating detailed analysis of the underlying flow physics. Although still idealised, span-periodic wings with large aspect ratios (e.g. in Fig.~\ref{fig:CRM_3D}) eliminate the influence of side-wall boundary conditions and enable the isolation and systematic investigation of both the 2D and 3D buffet instabilities in a controlled environment. Furthermore, simulations have the additional benefit of providing full access to data from the entire flow field, which is not possible in experiments.

Owing to the large aspect ratios required for the 3D instability to develop and coexist with the 2D mode \citep{LSH2024_narrow_buffet}, infinite-wing configurations to date have almost always been limited to low-fidelity RANS/URANS simulations \citep{PDL2020,DR2020} and GSA approaches \citep{CGS2019,PBDSR2019,PDBLS2021,HT2021}. \blue{However, the validity of these results is often questioned due to the limitations inherent to approximate turbulence models, which are known to be inaccurate/inconsistent in the presence of largely separated boundary layers and unsteadiness. These flow features are commonplace in off-design aerodynamics phenomena such as buffet and stall}. \blue{In practice, reproducing the low-frequency buffet unsteadiness with URANS often requires careful model selection, specific model corrections, and thorough sensitivity studies/tuning with respect to numerical settings, particularly for grid resolution/topology and temporal discretization (time step and sub-iterations). Also, the interplay between small-scale turbulence and three-dimensional buffet cell structures remains a critical, unresolved question.} High-fidelity approaches such as the implicit large eddy simulations in this work \citep{grinstein2007implicit} aim to overcome these limitations by solving the unsteady compressible Navier–Stokes equations directly without statistical averaging, modelling assumptions, or parameter tuning, thereby enabling a more accurate representation of the relevant unsteady flow physics and shock-induced boundary-layer separation \citep{grinstein2007implicit,RITOS_ILES_DNS}. Despite their clear advantages, the computational cost of LES/DNS simulations has naturally been limited to quasi-2D buffet \citep{FK2018,Moise2023_AIAAJ} on extremely small aspect ratios ($AR = L_z / c \sim 0.05$–$0.25$, where $L_z$ is the spanwise width and $c$ the airfoil chord), which are insufficient to capture fully-3D buffet structures of wavelength $\lambda_{z,3D} \sim 1$–$1.5c$, or, even, provide domain independent solutions of the 2D mode \citep{LSH2024_narrow_buffet}. 

Due to having access to large national scale HPC resources and the development of a computationally efficient and adaptable GPU-accelerated solver \citep{OpenSBLI_2021_CPC,LSSMSH_OpenSBLIv3_CPC2025}, we are able to perform extremely large scale-resolving ILES/DNS calculations on wide aspect ratio infinite wings. The flow solver generates a large number of samples of the time evolution of the flow fields which are then fed to data-driven modal analysis techniques, such as the spectral proper orthogonal decomposition (SPOD) \citep{towne_schmidt_colonius_2018}. This method is used to identify dominant frequencies and extract coherent flow structures of the most energetic modes and has already proven useful to determine the governing mechanisms of various fluid flow problems \citep{ModalDecomp_Taira2017}. 

Our recent numerical study on unswept ($\lambda = 0^{\circ}$) infinite wings \citep{lusher2025highfidelity_JFM} presented the first ILES at aspect ratios up to $AR=3$, demonstrating the natural emergence of strongly 3D buffet dynamics even in the absence of sweep or complex geometry. Due to the lack of imposed sweep, no preferential convection direction of the spanwise perturbations was observed. Building on this work, the present contribution performs a new set of simulations comprising of approximately $N \sim 8 \times 10^{9}$ mesh points (Fig.~\ref{fig:CRM_3D}) per case, with the objective of linking the \blue{intermittent quasi-stationary low-frequency ($St \sim 0.02$)} 3D mode identified in \citet{lusher2025highfidelity_JFM} to the intermediate-frequency ($St \sim 0.3$), spanwise-travelling 3D mode observed on swept wings in both low-fidelity simulations \citep{IR2015,ST2017,PDL2020,PDBLS2021,HT2021,SH2023} and experiments \citep{SKNNNA2018}. To this end, modal analysis techniques are applied to the resulting high-fidelity datasets in Section \S\ref{sec:SPOD}, enabling the isolation and comparison of the spatio-temporal energy content of the 2D and 3D modes. A representative range of sweep angles is considered, from zero to those commonly found on commercial aircraft ($\lambda \sim 35^{\circ}$).

\section{Computational method and problem specification}\label{sec:methods}

\subsection{Governing equations and numerical methods}\label{subsec:OpenSBLI}

For this transonic flow problem, the base governing equations are taken to be the non-dimensional compressible version of the Navier-Stokes equations for an ideal fluid. Applying conservation of mass, momentum, and energy, in the three spatial directions $x_i$ $\left(i=0, 1, 2\right)$, results in a system of five coupled partial differential equations to solve. These equations are defined for a density $\rho$, pressure $p$, temperature $T$, total energy $\rho E$, and velocity components $u_k$ as
\begin{align}\label{ns_eqn}
\frac{\partial \rho}{\partial t} &+ \frac{\partial}{\partial x_k} \left(\rho u_k \right) = 0,\\
\frac{\partial}{\partial t}\left(\rho u_i\right) &+ \frac{\partial}{\partial x_k} \left(\rho u_i u_k + p \delta_{ik} - \tau_{ik}\right) = 0,\\
\frac{\partial}{\partial t}\left(\rho E\right) &+ \frac{\partial}{\partial x_k} \left(\rho u_k \left(E + \frac{p}{\rho}\right) + q_k - u_i \tau_{ik}\right) = 0,
\end{align}
with heat flux $q_k$ and stress tensor $\tau_{ij}$ defined as
\begin{equation}\label{heat_flux}
q_k = \frac{-\mu}{\left(\gamma - 1\right) M_{\infty}^{2} Pr Re}\frac{\partial T}{\partial x_k}, \quad \tau_{ik} = \frac{\mu}{Re} \left(\frac{\partial u_i}{\partial x_k} + \frac{\partial u_k}{\partial x_i} - \frac{2}{3}\frac{\partial u_j}{\partial x_j} \delta_{ik}\right).
\end{equation}
\blue{$Pr=0.71$}, $Re=500,000$, and $\gamma=1.4$ are the Prandtl number, Reynolds number and ratio of specific heat capacities for an ideal gas, respectively. Support for curvilinear meshes is provided by using body-fitted meshes with a coordinate transformation. The equations are non-dimensionalised by a reference velocity, density and temperature $\left(U^{*}_{\infty}, \rho^{*}_{\infty}, T^{*}_{\infty}\right)$. In this work, the reference conditions are taken as the freestream quantities. For a reference freestream Mach number $M_{\infty}$, the pressure is defined as
\begin{equation}\label{pressure_eqn}
p = \left(\gamma - 1\right) \left(\rho E - \frac{1}{2} \rho u_i u_i\right) = \frac{1}{\gamma M^{2}_{\infty}} \rho T.
\end{equation}
Temperature dependent \blue{dynamic viscosity} is evaluated with Sutherland's law such that
\begin{equation}
\mu(T) = T^{\frac{3}{2}}\frac{1 + C_\textrm{Suth}}{T + C_\textrm{Suth}},
\end{equation}
with $C_\textrm{Suth} = T_{S}^{*} / T_{\infty}^{*}$, for $T_{S}^{*} = 110.4K$ and \red{reference temperature of $T_\infty^{*} = 273.15K$}. Skin-friction is defined for a wall shear stress $\tau_w$, as
\begin{equation}
C_f = \frac{\tau_w}{0.5 \rho_{\infty}^{*} U_{\infty}^{*2}}.
\end{equation}
The lift coefficient is evaluated over the airfoil surface with arc-length $s_{total}$ as
\begin{equation}
C_L = \frac{1}{0.5\rho_{\infty}^{*} U_{\infty}^{*2}}\int_{s=0}^{s=s_{total}} -S(p_w - p_{\infty}) \left|\cos(\theta)  \right| ds
\end{equation}
where $\theta$ is the inclination angle at the surface, and $p_w$ and $p_{\infty}$ are the wall and freestream pressures, respectively. The $S$ term represents the sign of the grid metrics depending on whether the coordinate around the airfoil is increasing/decreasing with respect to the grid index, to give the appropriate force contributions from the pressure/suction side of the airfoil. Quantities such as surface distributions of $C_f$ and $C_p$ in line plots are averaged in time and in the spanwise direction, unless otherwise stated. Aerodynamic forces such as $C_L$ and Power Spectral Density (PSD) plots of lift fluctuations are calculated at a single spanwise section ($z=L_z/2$) to detect the passing of spanwise propagating phenomena (i.e. buffet cells). Given the imposed periodicity of the problem, any spanwise section of the wing is equivalent to describe the buffet cells convective in the cross-flow direction. 

All implicit large eddy simulations \citep{grinstein2007implicit} in this work are performed in the OpenSBLI open source high-order finite-difference compressible flow solver \citep{OpenSBLI_2021_CPC,LSSMSH_OpenSBLIv3_CPC2025}. In addition to the verification and validation cases contained in the public code releases \citep{OpenSBLI_2021_CPC,LSSMSH_OpenSBLIv3_CPC2025}, OpenSBLI has also been cross-validated against six other independently-developed flow solvers using a range of diverse numerical methodologies by \citet{NATO2024_TGV}. OpenSBLI has a variety of numerical methods available to users. In the present work, fourth-order explicit central differences are used to compute spatial derivatives for both convective and diffusive terms. Convective terms are written in a cubic split form to boost numerical stability. A fifth-order Weighted Essentially Non-Oscillatory (WENO-Z) scheme is used at grid locations identified by a shock sensor. A detailed description of the numerical methods is given in \cite{LSSMSH_OpenSBLIv3_CPC2025}. Time-advancement is performed by an explicit fourth-order 5-stage low-storage Runge-Kutta scheme. Dispersion Relation Preserving (DRP) filters are applied to the freestream. 

Airfoil-specific applications of this code configuration include the laminar buffet simulations on the V2C airfoil profile in \citet{LZSH2023}, and turbulent buffet on the NASA-CRM airfoil geometry in \citet{LSH2024_narrow_buffet,LSSMSH_OpenSBLIv3_CPC2025,lusher2025highfidelity_JFM}. The turbulent cases included cross-validation against buffet predictions from unsteady-RANS and GSA (as developed in \cite{SH2023}). The wide-span unswept transonic buffet cases presented in \cite{lusher2025highfidelity_JFM} form the basis for the present study on sweep effects, as described next.

The large-scale simulations performed in this work were run simultaneously on both GPU and CPU high performance computing clusters. The total combined computational runtime cost was approximately 648,000 GPU-hours using NVIDIA V100 GPUs with CUDA+MPI (or,  equivalently, $3.8\times 10^8$ Fujitsu A64FX CPU-core-hours with hybrid OpenMP+MPI on CPU). Data storage for the study (time series of 2D snapshots, grid files, and restart files) totalled around 34TB of disk space.

\subsection{Transonic airfoil buffet problem description}\label{subsec:problem_description
}

The simulations are performed at a moderate Reynolds number of $Re_c=0.5\times10^6$ based on airfoil chord length, $c$, and a transonic freestream Mach number of $M_\infty=0.72$. These conditions are sufficient to obtain fully turbulent flow upstream of the SBLI \citep{LSH2024_narrow_buffet}. The calculations are performed on the same setup and mesh used in \citet{lusher2025highfidelity_JFM} for the largest aspect ratio considered of $AR=3$. The 65\% semi-span station of the NASA Common Research Model (CRM) wing is uniformly extruded in the periodic spanwise direction to $L_z = 3c$ (right plot in Fig.\ref{fig:CRM_3D}). The sharp trailing edge configuration of this geometry was used.

The numerical domain consists of one airfoil C-mesh connected to two wake blocks. The in-flow boundary is set at a distance of $25c$ with the outlet $5c$ downstream of the aerofoil. The resolutions in the azimuthal and normal to the surface directions are $\left(2249, 681\right)$ and $\left(701, 681\right)$ for airfoil and wake blocks, respectively. The domain has been discretised in the spanwise direction using 3000 equi-spaced points, for a total mesh count of around $N=7.46\times 10^9$. \red{A spanwise grid study at buffet conditions for the same CRM configuration was reported in \citet{LSH2024_narrow_buffet} on narrow domains $(L_z = 0.05c)$. Based on that assessment, a medium spanwise spacing of $\Delta z = 0.001$ was adopted for the present cases (as in previous unswept wide-span simulations \cite{lusher2025highfidelity_JFM}) to keep simulations with $AR = 3$ computationally feasible. Around the primary shock at $x\approx 0.4$ (Fig.~\ref{fig:Cp_Cf}), the near-wall resolution corresponds to $\left(\Delta x^{+}, \Delta y^{+}, \Delta z^{+}\right)=\left(6.1,2.2,14.8\right)$, while in the attached turbulent region downstream of the shock the maximum values at $x\approx0.7$ are $\left(\Delta x^{+}, \Delta y^{+}, \Delta z^{+}\right)=\left(3.9,1.1,7.6\right)$. These are within commonly used ranges for LES of turbulence ($\Delta x^{+} \approx 100$, $\Delta y^{+} \approx 1$, $\Delta z^{+} \approx 30$, \citet{davidson2009large}). In addition to the mesh sensitivity tests in the previous works, an extra sensitivity test (omitted here for brevity) was performed with a wall-refined mesh on a narrow domain with $\Delta y^{+} < 1$ everywhere, and found to be essentially identical to the baseline mesh in terms of mean $C_p$, $C_f$ and shock oscillation frequency}. \red{To ensure adequate resolution across the full boundary layer, only weak grid stretching is applied; the boundary layer is discretized with 80 and 195 points at $x=0.4$ and $x=0.7$, respectively}.

\red{To place the present study in the context of comparable LES of airfoil buffet in the literature (albeit, on narrow domains): the independent works of \citet{dandois_mary_brion_2018} and \citet{ZMS2022} both utilized 200 million grid points per $0.05c$ spanwise width of the computational domain at a chord-based Reynolds number of $3\times 10^{6}$. The present study utilizes 125 million points per $0.05c$ of spanwise width despite the Reynolds number being six times lower ($Re_c = 0.5 \times 10^6$). The recent airfoil buffet study of \citet{moise_zauner_sandham_2022} consisted of 75 million points at the same Reynolds number as here}. \red{These comparable scale-resolving buffet LES/ILES studies on structured grids used similar numerical methods to ours, demonstrating the suitability of the present mesh for resolving turbulent scales and the unsteady shock boundary layer interaction under buffet conditions}.

Turbulent flow conditions are obtained by forcing a time-dependent blowing/suction tripping strip imposed at $x=0.1c$ on both sides of the airfoil (right plot in Fig.\ref{fig:CRM_3D}). For more detailed information, the reader is referred to \citet{LSH2024_narrow_buffet,lusher2025highfidelity_JFM}, where cross-validations against URANS and GSA from a separate flow solver are also reported.

In our previous work \citep{lusher2025highfidelity_JFM}, only unswept wings were considered and nominal freestream Mach and Reynolds numbers (based on 2D chord length, $c = 1$) of $M_{\infty,2D} = 0.72$ and $Re_{\infty, 2D} = 0.5\times 10^6$ were imposed. Given the periodicity at the spanwise boundaries in the present study, we can simulate a swept wing by imposing a spanwise velocity component $w=\tan \lambda$ to the inflow, rotating the freestream velocity vector by a sweep angle $\lambda$. Following \citet{PDL2020}, Mach and Reynolds numbers are thus modified as $M_{\infty} = M_{\infty,2D}  / \cos(\lambda)$ and $Re_{\infty} = Re_{\infty,2D}  / \cos(\lambda)^2$. Since the angle of attack is applied by rotating the geometry, there is no need to modify the incidence of the incoming flow for different sweep angles. Different sweep strengths are imposed by selecting $\lambda = \left[0^{\circ}, 5^{\circ}, 10^{\circ}, 15^{\circ}, 25^{\circ}, 35^{\circ}\right]$.

Our recent results \citep{lusher2025highfidelity_JFM} demonstrated that by varying the angle of attack, $\alpha$, essentially-2D buffet (at $\alpha = 5^{\circ}$) and 3D buffet (at $\alpha = 6^{\circ}$) behaviour can be observed. This was true even when the spanwise width was sufficiently large ($L_z = 2-3c$) to accommodate the 3D wavelength $(\lambda_z \sim 1-1.5c)$ (e.g. $\lambda_z=1c$ was found by \citet{HT2021}). The higher angle of attack leads to more extensive mean flow separation on the upper surface of the airfoil, which was a necessary condition to observe the spanwise 3D separation bubbles. To collect statistically meaningful quantities, multiple low-frequency 2D shock buffet cycles are simulated in each case to observe both low- and intermediate-frequency phenomena and enable the data-driven modal analysis that forms the focal point of this study.

\subsection{Data-driven modal analysis tools}\label{subsec:SPOD_VMD_description}

To analyse the spatio-temporal modes present in the high-dimensional flow database, an SPOD method \citep{towne_schmidt_colonius_2018} is applied to the 2D flow snapshots. We apply the open source PySPOD library \citep{mengaldo2021pyspod,pyspod_2024} to perform the analysis. The SPOD implementation and analysis of OpenSBLI simulation data has been demonstrated and cross-validated by \citet{HLS2023_IJHFF_SPOD} and \citet{LSSMSH_OpenSBLIv3_CPC2025}. In the present work, off-wall suction-side snapshots in the $x$-$z$ plane are processed at a sampling period of $\Delta T_{SPOD,sampling} = 1000\Delta t$, for a non-dimensional time-step of $\Delta t = 5\times 10^{-5}$. \red{The flow data is sampled either on the airfoil surface (wall pressure), or off-wall at a height of $1.5\times 10^{-4}c$ (for velocity-based SPOD)}. The selected collection interval provides an upper frequency resolution of $St = 10$, which is sufficient for the phenomena we investigate (typically, 2D buffet: $St \leq 0.1$, intermediate separation bubble modes: $St \sim 0.5$, and vortex-shedding/wake modes: $St \sim 1 - 2$) \citep{Moise2023_AIAAJ}.

By default, the data sets are divided into three segments with 50\% overlap. To enforce periodicity in each segment, a Hanning function is applied. \red{Each of these segments contains 30 convective time units, corresponding to approximately 7 periods of the 3D buffet cell mode (at $\lambda=25^{\circ}$, Fig.~\ref{fig:SPOD_modes_AoA6_shock_and_cell})}. \red{When presenting the main results}, the SPOD eigenvalue frequency spectra are plotted only for the first SPOD mode and compared with the PSD of the lift-coefficient fluctuations. \red{Sub-dominant SPOD modes were also assessed but found to provide minimal additional insight}. \red{To assess and quantify the sensitivity of the SPOD modes to the number of Welch segments, mode‑shape agreement across different numbers of segmentations is performed using a normalized W-inner‑product convergence criteria between SPOD modes at a given frequency following \citet{blanco2022improved}, such that}
\red{
\begin{equation}\label{eqn:SPOD_convergence}
\mathcal{O}_{W}(\boldsymbol{\phi(\omega)},\boldsymbol{\psi(\omega)})
=\frac{\bigl|\boldsymbol{\phi(\omega)}^{H}\mathbf{W}\,\boldsymbol{\psi(\omega)}\bigr|}
{\sqrt{\boldsymbol{\phi(\omega)}^{H}\mathbf{W}\,\boldsymbol{\phi(\omega)}}\;
 \sqrt{\boldsymbol{\psi(\omega)}^{H}\mathbf{W}\,\boldsymbol{\psi(\omega)}}}
,\qquad
E_{W}=1-\mathcal{O}_{W}^{2}.
\end{equation}
}\red{Here, $\boldsymbol{W}$ are the SPOD weights, and $\phi(\omega)$ and $\psi(\omega)$ are the SPOD modes computed with different segment lengths, evaluated at a selected single frequency $\omega$. The quantity $E_{W}$ is a measure of the error, where $\mathcal{O}_{W}$ represents perfect alignment. Values of $\mathcal{O}_{W} \geq 0.95$ are typically considered to be a suitable threshold to demonstrate excellent convergence of the modes \citep{abreu2021spanwise}}. For visualization purposes, real and imaginary parts of the SPOD modes have been used to reconstruct the mode temporal evolution. SPOD decompositions are presented based on the pressure field, streamwise $u$-velocity, and spanwise $w$-velocity components, as these proved to be the most illustrative for our purposes. The cell volumes are used for the SPOD weights. For brevity, side ($x$-$y$)-view SPOD modes are not shown, as the focus in this work is on the spanwise convection of 3D perturbations and dynamics within the $x$-$z$ plane. SPOD modes on the same airfoil geometry in the $x$-$y$ plane can be found in \citet{lusher2025highfidelity_JFM}.

\section{Results}\label{sec:results}

\subsection{Recap on quasi-2D and 3D buffet characteristics on unswept wide-span infinite wings}

Our recent transonic buffet studies on both narrow- ($AR = 0.05-0.5$) \cite{LSH2024_narrow_buffet} and wide-span ($AR = 1-3$) \cite{lusher2025highfidelity_JFM} aspect ratio ($AR$) infinite wings concentrated on unswept ($\lambda = 0^{\circ}$) configurations of the  NASA-CRM airfoil at different aspect ratios and angles of attack ($\alpha$). Note, the use of `narrow' and `wide' here are in the context of what is realistically feasible for LES/DNS studies with current compute capability, and not a comment on the 3D mode wavelength itself. Low-fidelity stability-based methods and URANS of infinite wings (for example, \citet{PDL2020}, and \citet{HT2021}), often go up to $AR=6-10$. For the low aspect ratio wings ($AR \sim 0.1$) in \cite{LSH2024_narrow_buffet}, the onset of the quasi-2D low-frequency buffet mode ($St \approx 0.08$) was found to occur at these conditions at an angle of attack of $\alpha = 4.5^{\circ}$. Good agreement was observed for both onset criteria and buffet frequency predictions between ILES, URANS, and GSA \cite{LSH2024_narrow_buffet}. Domain independence of the quasi-2D buffet mode was demonstrated for the scale-resolving and RANS simulations, which remained essentially 2D at $\alpha =5^{\circ}$ up to $AR=0.5$ with no spanwise 3D structures present.

\begin{figure}
\centering
\includegraphics[width=1\columnwidth]{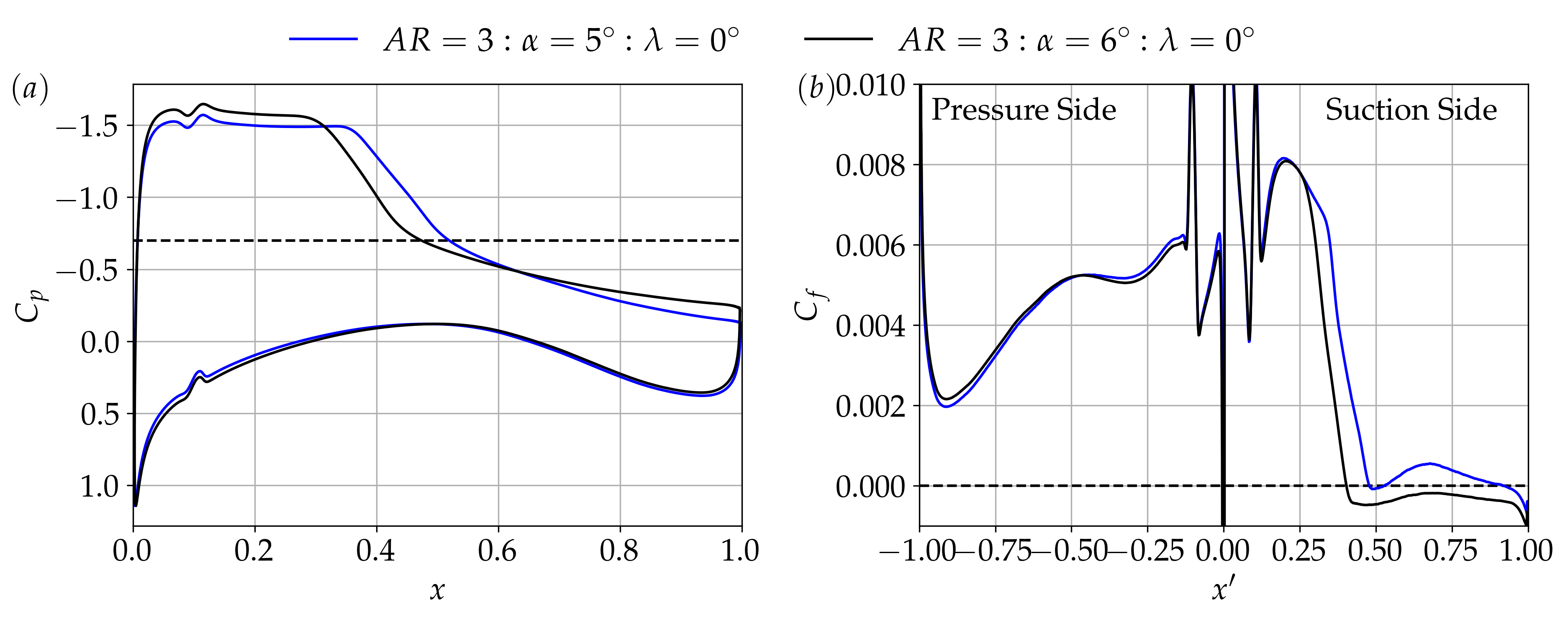}
\caption{Span- and time-averaged mean $(a)$ pressure coefficient and $(b)$ skin-friction, for two unswept $\lambda = 0^{\circ}$ cases corresponding to (blue, $\alpha=5^{\circ}$) minimally-, and (black, $\alpha=6^{\circ}$) largely-separated flow at the SBLI $(x \approx 0.4-0.5)$.}
\label{fig:Cp_Cf}
\end{figure}

This work was later extended in \cite{lusher2025highfidelity_JFM} to consider ILES of wide-span aspect ratios up to $AR=3$, over an active buffet range of angles of attack of $\alpha = \left[5^{\circ},6^{\circ},7^{\circ}\right]$. The analysis continued the focus on unswept $(\lambda = 0^{\circ})$ airfoil configurations. At $AR=1$ and above, large 3D spanwise separation cells were observed in cases where significant time-averaged flow separation was present $(\alpha \geq 6^{\circ})$. The spanwise wavelength of the 3D mode was $\lambda_z \sim 1-1.5c$ airfoil chords, dependent on the aspect ratio and angle of attack considered. At those unswept conditions, a \blue{quasi-stationary} 3D mode was observed at very low frequencies ($St\approx 0.02$), below the characteristic frequency of the chordwise shock oscillation component ($St = 0.08$). The separation cells were found to be intermittent, strengthening during low-lift phases of the buffet cycle as the shock moved upstream in the direction of the leading edge of the airfoil. The number of separation cells also varied between one and two, at different points of the multiple periods of the low-frequency cycle. The three-dimensionality was persistent over numerous cycles and not a transient effect.

While previous stability-based buffet studies around RANS base flows \cite{CGS2019} on straight wings have predicted simultaneous onset of the 2D and 3D buffet modes, the ILES in \cite{lusher2025highfidelity_JFM} presented cases above buffet onset ($\alpha = 5^{\circ}$) that remained essentially-2D with a uniform shock across the entire span-width. While this could also be due to the different geometries used (OAT15A vs CRM), a clear chordwise low-frequency shock unsteadiness was present, despite the absence of any 3D separation cells. SPOD analysis of the spanwise velocity component, $w$, showed a very weak spanwise modulation in the otherwise 2D flow field.

\blue{\subsection{Overview of the cases considered in the present study}}

As detailed in the case overview listing in Table~\ref{tab:tab1}, these two buffet configurations ($\alpha = 5^{\circ}$ and $6^{\circ}$) are selected to investigate the effect of cross flow on the 2D and 3D buffet modes. A range of sweep angles are considered between $\lambda = \left[0^{\circ}, 35^{\circ}\right]$, depending on the angle of attack. For the main results, the airfoil aspect ratio is fixed to $AR=3$, to accommodate two wavelengths of the expected 3D mode.

\begin{table}[b!]
  \caption{List of airfoil buffet cases considered and their numerical parameters.}\label{tab:tab1}
  \begin{ruledtabular}
    \begin{tabular}{c c c c c c}
    \textbf{Case Name} & \textbf{$M_{\infty,2D}$} & \textbf{$Re_{c,\infty,2D}$} & \textbf{Aspect Ratio ($AR = L_z / c$)} & \textbf{Angle of Attack ($\alpha^{\circ}$)} & \textbf{Sweep Angle ($\lambda^{\circ}$)} \\
    \hline
    AR2-AoA5-Lambda0 & 0.72 & $0.5\times10^6$ & 2 & 5 & 0 \\
    AR3-AoA5-Lambda0 & 0.72 & $0.5\times10^6$ & 3 & 5 & 0 \\
    AR3-AoA5-Lambda5 & 0.72 & $0.5\times10^6$ & 3 & 5 & 5 \\
    AR3-AoA5-Lambda25 & 0.72 & $0.5\times10^6$ & 3 & 5 & 25 \\
    AR3-AoA6-Lambda0 & 0.72 & $0.5\times10^6$ & 3 & 6 & 0 \\
    AR3-AoA6-Lambda5 & 0.72 & $0.5\times10^6$ & 3 & 6 & 5 \\
    AR3-AoA6-Lambda10 & 0.72 & $0.5\times10^6$ & 3 & 6 & 10 \\
    AR3-AoA6-Lambda15 & 0.72 & $0.5\times10^6$ & 3 & 6 & 15 \\
    AR3-AoA6-Lambda25 & 0.72 & $0.5\times10^6$ & 3 & 6 & 25 \\
    AR3-AoA6-Lambda35 & 0.72 & $0.5\times10^6$ & 3  & 6 & 35 \\
    \end{tabular}
  \end{ruledtabular}
\end{table}

Span- and time-averaged profiles of $(a)$ pressure coefficient and $(b)$ skin-friction are reported in Fig.\ref{fig:Cp_Cf} for the two angles of attack at $\lambda=0^{\circ}$. The trip location is visible at $x=0.1c$ on both sides of the airfoil (Fig.\ref{fig:CRM_3D}). The increase in angle of attack shifts the mean shock wave position farther upstream on the upper surface, whereas the pressure side of the airfoil is seen to be less sensitive to the change. The time-averaged pressure gradient imposed by the shock \orange{(Fig.\ref{fig:Cp_Cf}, around $0.3 < x < 0.55$)} is spread over a wide region ($\sim25\%c$) of the airfoil chord due to the chordwise motion associated with shock buffet \cite{Giannelis_buffet_review}.

The main physical difference between the two cases is observed in the skin-friction distribution around the mean shock location ($x \approx 0.4-0.5$). At $\alpha=5^{\circ}$, the flow is attached in a time/span-averaged mean sense, but at $\alpha=6^{\circ}$, a large separation bubble is present at the shock that extends downstream towards the trailing edge. Although the $\alpha=5^{\circ}$ case is attached in the mean, at low-lift phases of the buffet cycle it is still massively-separated instantaneously, downstream of the SBLI \cite{LSH2024_narrow_buffet,lusher2025highfidelity_JFM,LSSMSH_OpenSBLIv3_CPC2025}, as will be shown. The focus of this work is the buffet dynamics of these two configurations (minimally- ($\alpha = 5^{\circ}$) and largely-separated ($\alpha = 6^{\circ}$) in the mean) when sweep is introduced. Of particular interest are the effect of sweep on the \blue{intermittent quasi-stationary} unswept 3D separation cell mode ($St \approx 0.02$) and the relation between 3D buffet mode frequency and sweep angle as disturbances convect in the spanwise direction.\\

\subsection{Swept infinite wing transonic buffet: weak mean flow separation at the SBLI $(\alpha = 5^{\circ})$}\label{subsec:AoA5}
\begin{figure}
\centering
\includegraphics[width=1\columnwidth]{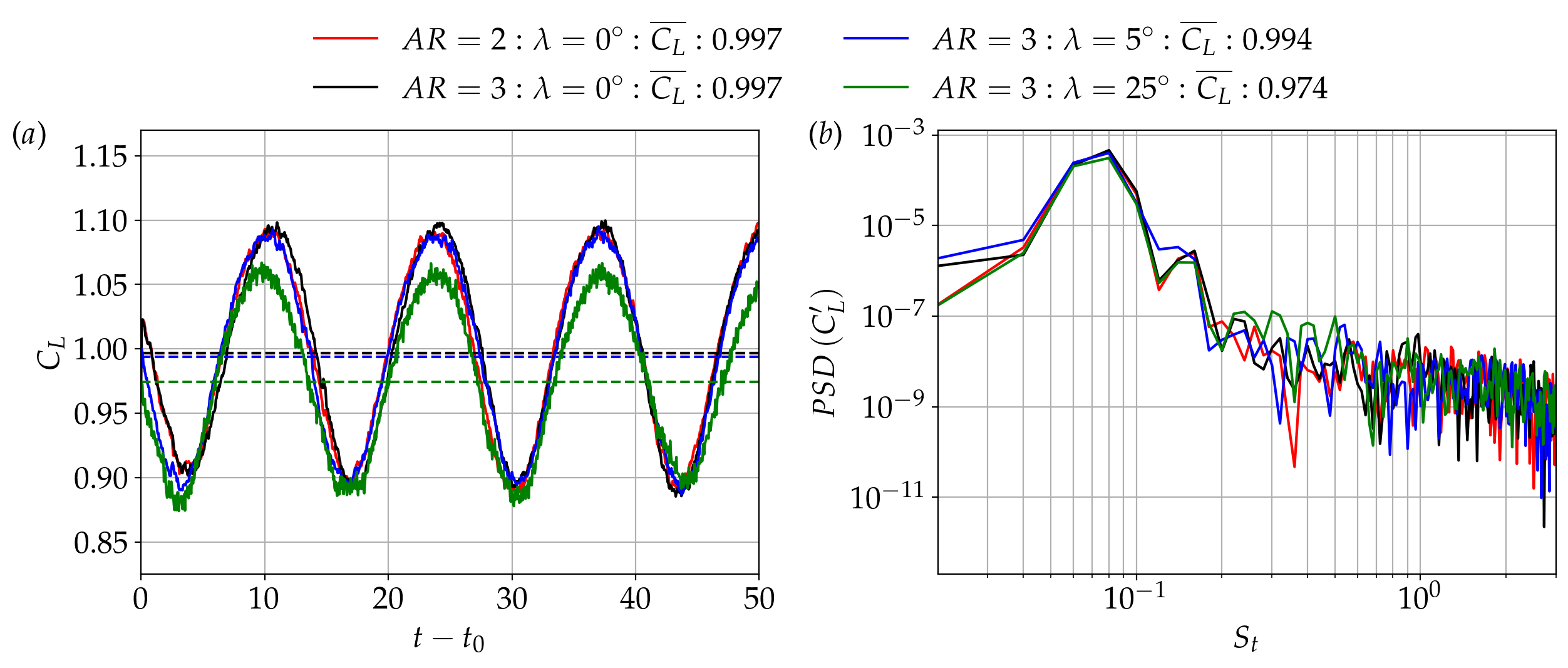}
\caption{Sectional evaluation at $z = L_z / 2$ of the $(a)$ lift coefficient and $(b)$ power spectral density of lift fluctuations. Showing the low-frequency unsteadiness in $\alpha=5^{\circ}$ cases at different aspect ratios and sweep angles.}
\label{fig:deg5_lines}
\end{figure}

Figure~\ref{fig:deg5_lines} reports sectional evaluation of the $(a)$ lift coefficient history and $(b)$ Power Spectral Density (PSD) of lift fluctuations at a single spanwise grid location ($z = L_z / 2$) for the $\alpha = 5^{\circ}$ case. Sweep angles of $\lambda = \left[0^{\circ}, 5^{\circ}, 25^{\circ}\right]$ are shown. Sectional evaluation is necessary here to assess macroscopic spanwise inhomogeneity which would be lost when performing a spanwise average (as in Fig.~\ref{fig:Cp_Cf}) \cite{PDL2020,PDBLS2021}. If large-scale 3D structures are present, a dependence on the $z$-location used for the evaluation would be obvious. The profiles at $AR=2$ and $AR=3$ for unswept cases overlap exactly. This reaffirms the span length independence of this $\alpha=5^{\circ}$ case at $AR=2$ from \cite{lusher2025highfidelity_JFM}, now up to $AR=3$. At these flow conditions, no \blue{macroscopic three-dimensional flow structures} are observed and the flow behaves in a quasi-2D manner with no aspect ratio dependence (at least up to $AR=3$). Identical prediction of the low-frequency 2D shock oscillation peak at $St = 0.08$ is noted in Fig~\ref{fig:deg5_lines}($b$). Introducing a small amount of sweep at $\lambda = 5^{\circ}$ has minimal effect on this case. The time-averaged $C_L$ decreases by only 0.3$\%$ and the low-frequency buffet mode is unchanged.
 
At the higher sweep angle of $\lambda=25^{\circ}$, the mean lift reduces by around 2\% relative to unswept conditions. Plante et al. \cite{PDL2020,PDBLS2021} reported a similar reduction in $C_L$ between the lift histories from swept/unswept unsteady-RANS cases. Despite the observed reduction in lift oscillation amplitude, the sectional evaluation shows the flow at $\lambda=25^{\circ}$ is still dominated by the chordwise shock oscillations, and behaves in a quasi-2D manner with no significant three-dimensionality (i.e. buffet cells). As seen in the PSD in Fig.~\ref{fig:deg5_lines}($b$), the swept $\lambda=25^{\circ}$ case still shows an identical low-frequency peak at $St=0.08$.

\begin{figure}
\centering
\includegraphics[width=1\columnwidth]{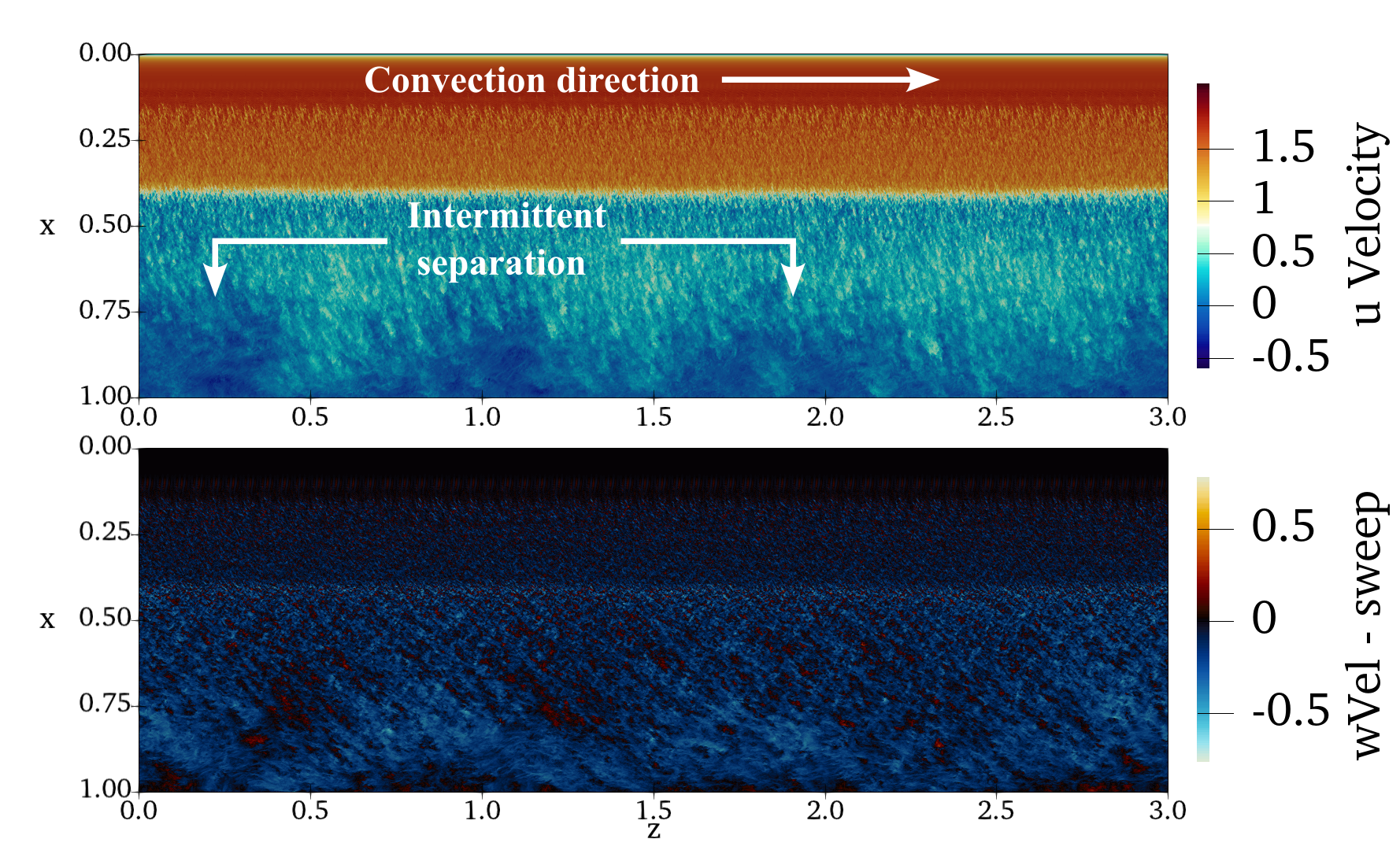}
\caption{Instantaneous flow visualisations over the suction side of the airfoil at $\lambda = 25^{\circ}$ and $\alpha = 5^{\circ}$. The top/bottom plots show the off-wall streamwise/spanwise velocity at $y=0.003c$. The shock front is observed to be quasi-2D across the span for all times at $\alpha=5^{\circ}$, despite the low-frequency buffet unsteadiness and intermittent separation patches near the trailing edge.}
\label{fig:AoA5_surface_contours}
\end{figure}

There is, however, minor additional energy content present in an intermediate Strouhal number range of $0.2 < St < 0.4 $ (green curve), roughly where buffet cells are expected to appear \citep{PDL2020}. Inspection of the off-wall flow showed that, at all times in the flow database, the shocked region of the flow (where buffet cells are expected \cite{Giannelis_buffet_review}) remains essentially-2D across the span, without any buffet cells being present along the shock. An example instantaneous visualization of the near-wall flow is given in the streamwise/spanwise-velocity contours in Fig.\ref{fig:AoA5_surface_contours} at $y=0.003c$ on the suction side of the airfoil. The contour levels have been centred in the bottom panel to remove the $w=\tan \left(25^{\circ}\right)$ sweep component. The position of the shock front is parallel to the leading edge of the airfoil at all spanwise locations, independent of the span width (up to $AR=3$). This was found to be true at all phases of the lift cycle (Fig.~\ref{fig:deg5_lines}). However, while the shock front remains essentially-2D across the span, a series of intermittently-present separation bubbles are observed across the spanwise width (dark blue), located in the reverse flow region adjacent to the trailing edge. These were observed to develop most prominently during low-lift phases as the boundary layer downstream of the SBLI periodically separates. The spanwise velocity component in the bottom panel shows no three-dimensionality at the shock beyond small-scale turbulent structures. At this angle of attack, while minor intermittent separation cells (akin to stall cells, as suggested by \citep{PDL2020,PDBLS2021}) are observed near the airfoil trailing edge, the SBLI remains quasi-2D across the span at all times, with no large-scale 3D unsteadiness.

To conclude this section, at this angle of attack of $\alpha=5^{\circ}$, the 2D chordwise shock buffet mode $(St \approx 0.08)$ is dominant, without any spanwise three-dimensionality at the shock. Regardless of whether the flow is unswept $(\lambda=0^{\circ})$ or swept $(\lambda=5^{\circ}, 25^{\circ})$, the lift oscillations and associated buffet characteristics are quasi-2D in nature (Fig.~\ref{fig:deg5_lines}), independent of sweep angle. Weak three-dimensionality is observed in the separated region downstream of the SBLI at $\lambda=25^{\circ}$, near to the trailing edge, and not directly coupled to the main shock wave. This case is investigated in more detail in the spectral mode decomposition analysis in Section \S\ref{sec:SPOD}.

\subsection{Swept infinite wing transonic buffet: strong mean flow separation at the SBLI $(\alpha = 6^{\circ})$}\label{subsec:AoA6}

\begin{figure}
\centering
\includegraphics[width=1\columnwidth]{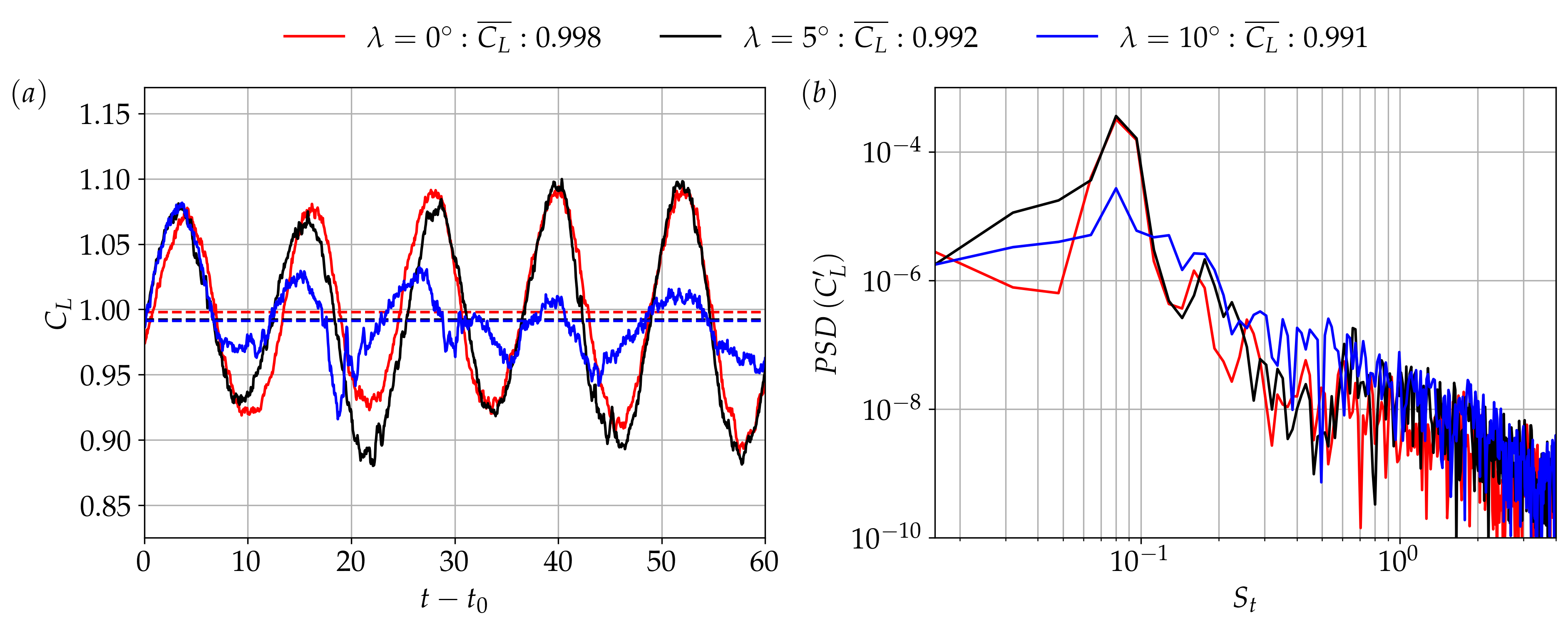}
\includegraphics[width=1\columnwidth]{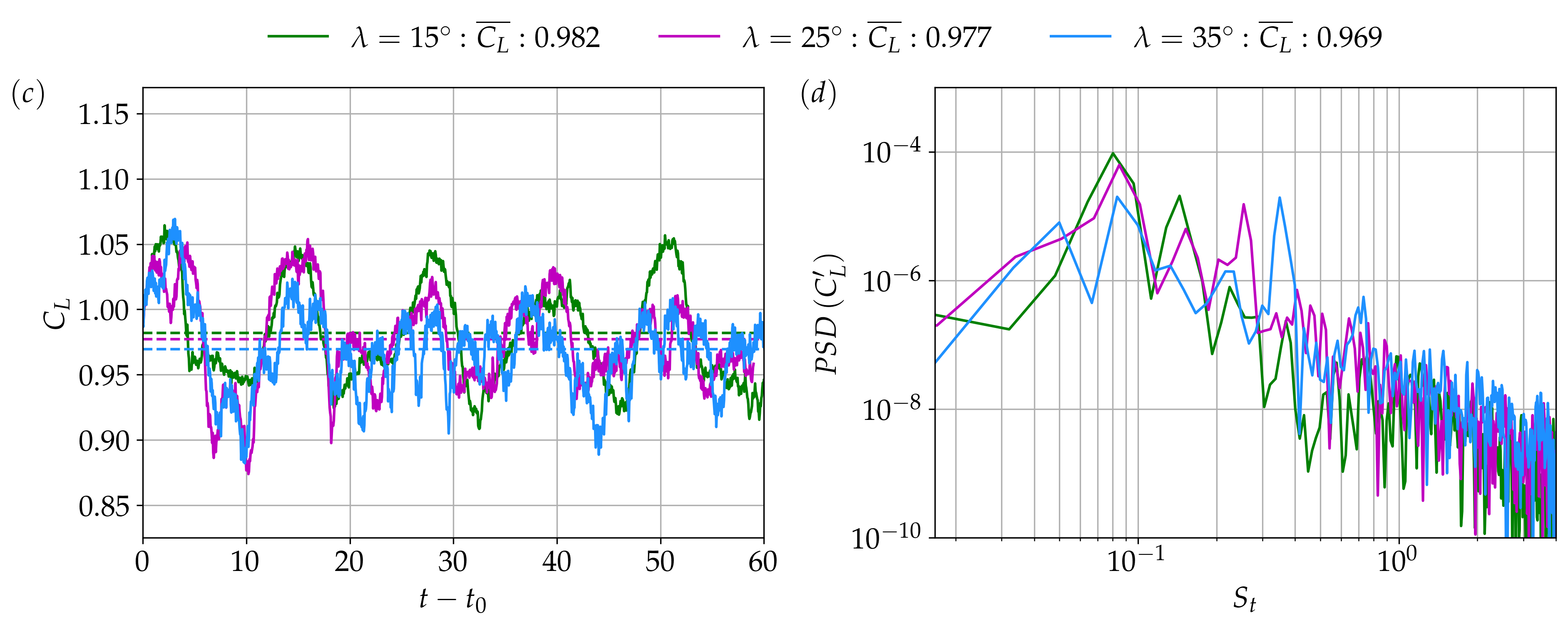}
\caption{Sectional evaluation of $(a,c)$ lift coefficient and $(b,d)$ PSD of lift fluctuations. Showing $\alpha=6^{\circ}$ buffet cases at $(a,b)$ zero/low- and $(c,d)$ high-sweep angles, demonstrating the monotonic frequency increase of the 3D buffet cell mode with sweep.}
\label{fig:deg6_lines_low}
\end{figure}

Results are presented in this section at the higher angle of attack of $\alpha= 6^{\circ}$, at which increased mean flow separation is present at the SBLI (Fig.~\ref{fig:Cp_Cf}($b$)). Based on our previous work on unswept configurations \cite{lusher2025highfidelity_JFM}, 3D buffet is expected at this angle of attack for the same flow conditions. Figure~\ref{fig:deg6_lines_low} compares the $(a,c)$ lift coefficient and $(b,d)$ PSD of lift fluctuations at a range of sweep angles between $\lambda=\left[0^{\circ}, 35^{\circ}\right]$. Sectional evaluation is again performed along a single section ($z = L_z / 2$) to detect deviations from quasi-2D behaviour, if present.

At $\lambda=5^{\circ}$, the influence of sweep is again found to be minor; the lift signal and the low-frequency peak in the PSD overlap almost exactly with the unswept case. The low-frequency unsteadiness appears as a well-resolved peak at $St = 0.08$ in both cases. As in the cases at $\alpha=5^{\circ}$ in the previous section, we are still within the Mach and angle of attack envelope of active buffet for this configuration \citep{LSH2024_narrow_buffet}. Below the frequency of the shock mode ($St=0.08$), the $\lambda=5^{\circ}$ case shows an increase in energy content around $0.03 < St < 0.05$, suggesting that the \blue{quasi-stationary} separation cell mode ($St\approx 0.02$) identified in \cite{lusher2025highfidelity_JFM} may be active and shifting in frequency with the introduction of sweep. This aspect is further examined by SPOD analysis in Section~\ref{sec:SPOD}.

As the sweep angle is further increased to $\lambda=10^{\circ}$, there is a notable suppression of the buffet oscillation amplitude. The lift history also becomes more irregular on a period-to-period basis, in contrast to the swept results at the lower angle of attack of $\alpha = 5^{\circ}$ (Fig.~\ref{fig:deg5_lines}($a$)). In the PSD of lift fluctuations (Fig.~\ref{fig:deg6_lines_low}($b$)), the low-frequency peak reduces in intensity compared to $\lambda=0^{\circ},5^{\circ}$. Additional energy content is also present between $0.1 < St < 0.2$, for which the SPOD analysis in the next section is required to extract more information from.

As we increased to higher sweep angles ($\lambda = 15^{\circ} - 35^{\circ}$) in Fig~\ref{fig:deg6_lines_low}$(c,d)$, although a low-frequency 2D mode is still present, the increasing cross-flow strength results in additional higher-frequency noise in the lift signal and greater irregularity between successive periods. Turning to the PSD of lift-fluctuations in Fig.~\ref{fig:deg6_lines_low}$(d)$, the low-frequency buffet mode at $St = 0.08$ is still captured. Its frequency is found to be independent of sweep angle. However, in contrast to the weakly-swept cases ($\lambda = 5^{\circ} - 10^{\circ}$), strong secondary peaks are now observed between $St \sim 0.15-0.35$, in the range expected for the 3D buffet cell mode \citep{Giannelis_buffet_review,PDL2020,VMD_Ohmichi2024}. The 3D mode here is observed to increase monotonically in Strouhal number as the imposed sweep increases $\left(\lambda, St\right) \approx \left(15^{\circ}, 0.16\right), \left(25^{\circ}, 0.25\right), \left(35^{\circ}, 0.33\right)$, in good agreement with trends shown by the low-fidelity (URANS) simulations of buffet cells on infinite swept wings in \cite{PDL2020,PDBLS2021} (albeit, at a higher Reynolds number and on a different airfoil profile to here).

\begin{figure}
\centering
\includegraphics[width=1\columnwidth]{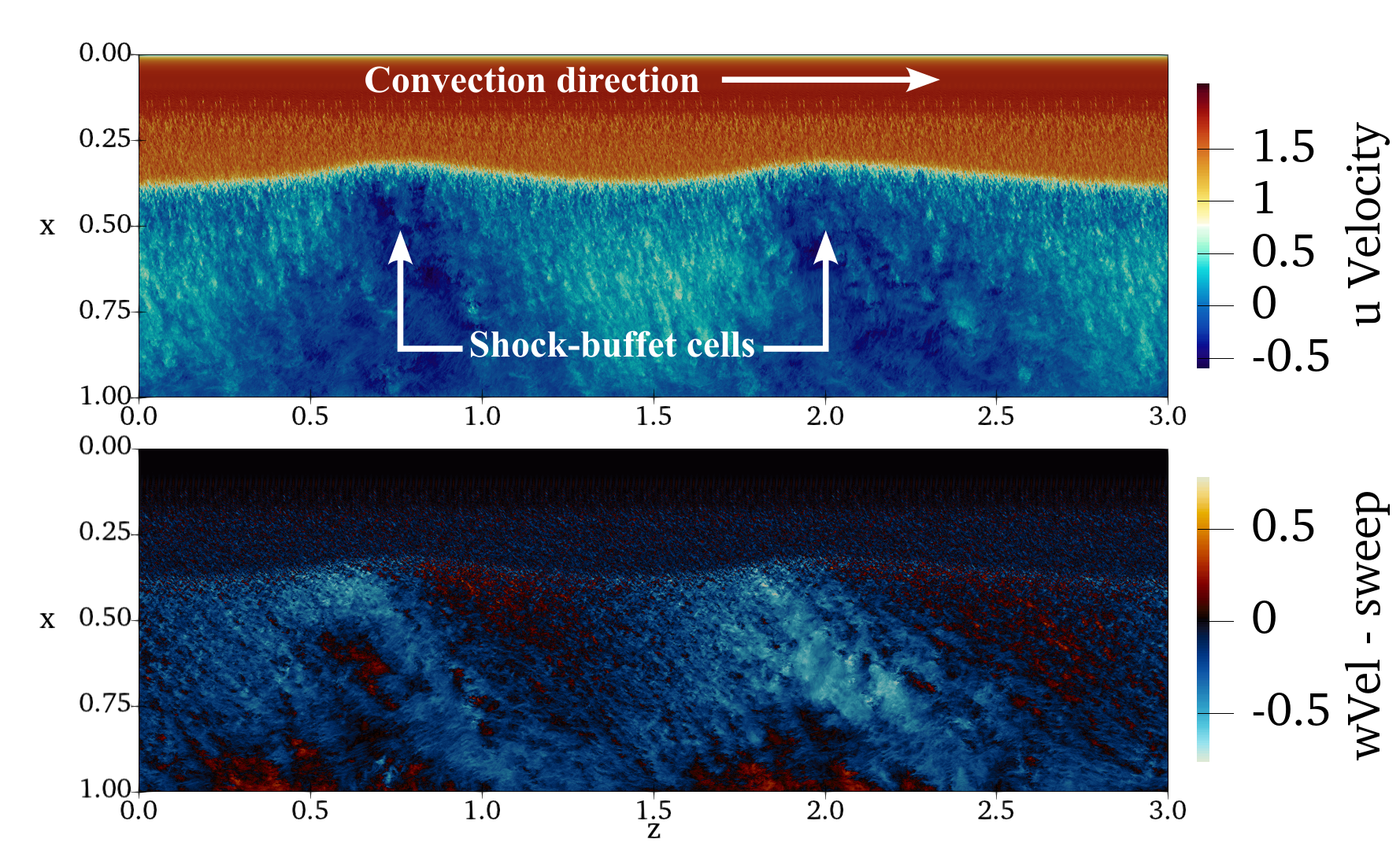}
\caption{Instantaneous flow visualisations over the suction side of the airfoil surface at a sweep angle of $\lambda = 25^{\circ}$, showing the 3D buffet cell instability. The top/bottom plots show the off-wall streamwise/spanwise velocity. buffet cells are highlighted.}
\label{fig:buffet_cells}
\end{figure}

Figure~\ref{fig:buffet_cells} visualizes the instantaneous flow field near the surface of the suction side of the airfoil at $\lambda = 25^{\circ}$. As before, the upper and lower panels show streamwise velocity and spanwise velocity minus the sweep component. Two large buffet cells are identified by the dark blue regions in the $u$-velocity, showing the local flow-reversal within the separation bubbles. The buffet cells have a spanwise wavelength of $\lambda_z\sim1-1.5c$ airfoil chords. The $w$-velocity exhibits alternating left- and right-moving (red/blue) fluid at the saddle point of the bubble along the separation line. At this higher angle of attack, \orange{the coexistence of 2D and 3D modes results in a complex shock topology. While the shock maintains a global chordwise motion consistent with the characteristic low-frequency buffet signature, it also exhibits spanwise periodic distortions driven by the emergence of higher-frequency 3D buffet cells}. Once sweep is imposed, the buffet cells convect across the span, which can be inferred from the left-to-right diagonal traces in the $w$-velocity. \orange{While the disturbances convect in the spanwise direction parallel to the leading edge, as the wing is swept, the 3D mode is oblique, containing components of motion in both the $x$ and $z$ directions}. Although the 2D shock-oscillation mode is active at both angles of attack ($\alpha=5^{\circ},6^{\circ}$), the presence of significant flow separation at the shock location at $\alpha=6^{\circ}$ is found to be a necessary condition for the 3D mode to be active in this region.

\section{Modal analysis of turbulent transonic buffet cells}\label{sec:SPOD}

A spectral proper orthogonal decomposition (SPOD) analysis method is applied to investigate the spectral energy content present in the near-wall boundary layer flow on the suction side of the airfoil for the flow configurations in the previous section. The aim is to identify the dominant 2D- and 3D-modes acting in transonic airfoil buffet and their characteristic frequencies, under the influence of increasing cross-flow. 

In each case, the SPOD algorithm is applied separately to three different flow variables that were found to be the most instructive. Namely, modes are examined in the streamwise $u$- and spanwise $w$-velocity components on the first off-wall point, and the surface pressure, $p$. The value in viewing decompositions of different flow variables will become apparent. Decompositions based on other flow variables (e.g. density, Mach number, $v$-velocity, and velocity aligned with the inflow, $U_\parallel = u\cos\lambda + w\sin\lambda$) were also examined, but were ultimately found to be less useful for our purpose of identifying 2D- and 3D-modes. Direct comparison is first made between the SPOD eigenvalue spectra and the PSD obtained from lift-fluctuations (Fig.~\ref{fig:deg5_lines}$(b)$, Fig.~\ref{fig:deg6_lines_low}$(b)$). Spatial modes are then plotted over the airfoil surface at the dominant frequencies.

\subsection{Modal analysis of the quasi-2D ($\alpha = 5^{\circ}$) buffet interaction}\label{subsec:AoA-5_SPOD}

\begin{figure}
\begin{center}
\includegraphics[width=0.495\columnwidth]{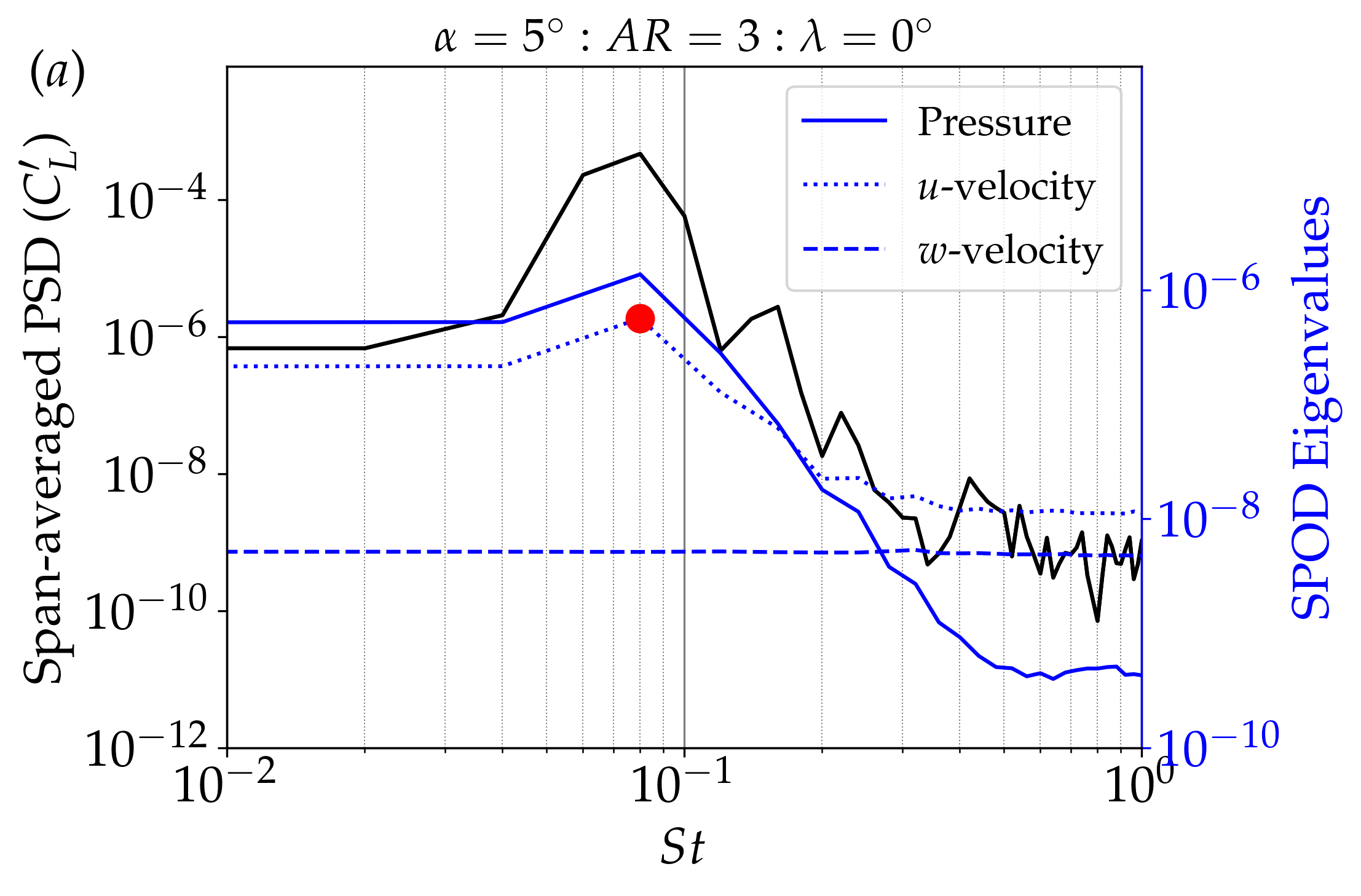}
\includegraphics[width=0.495\columnwidth]{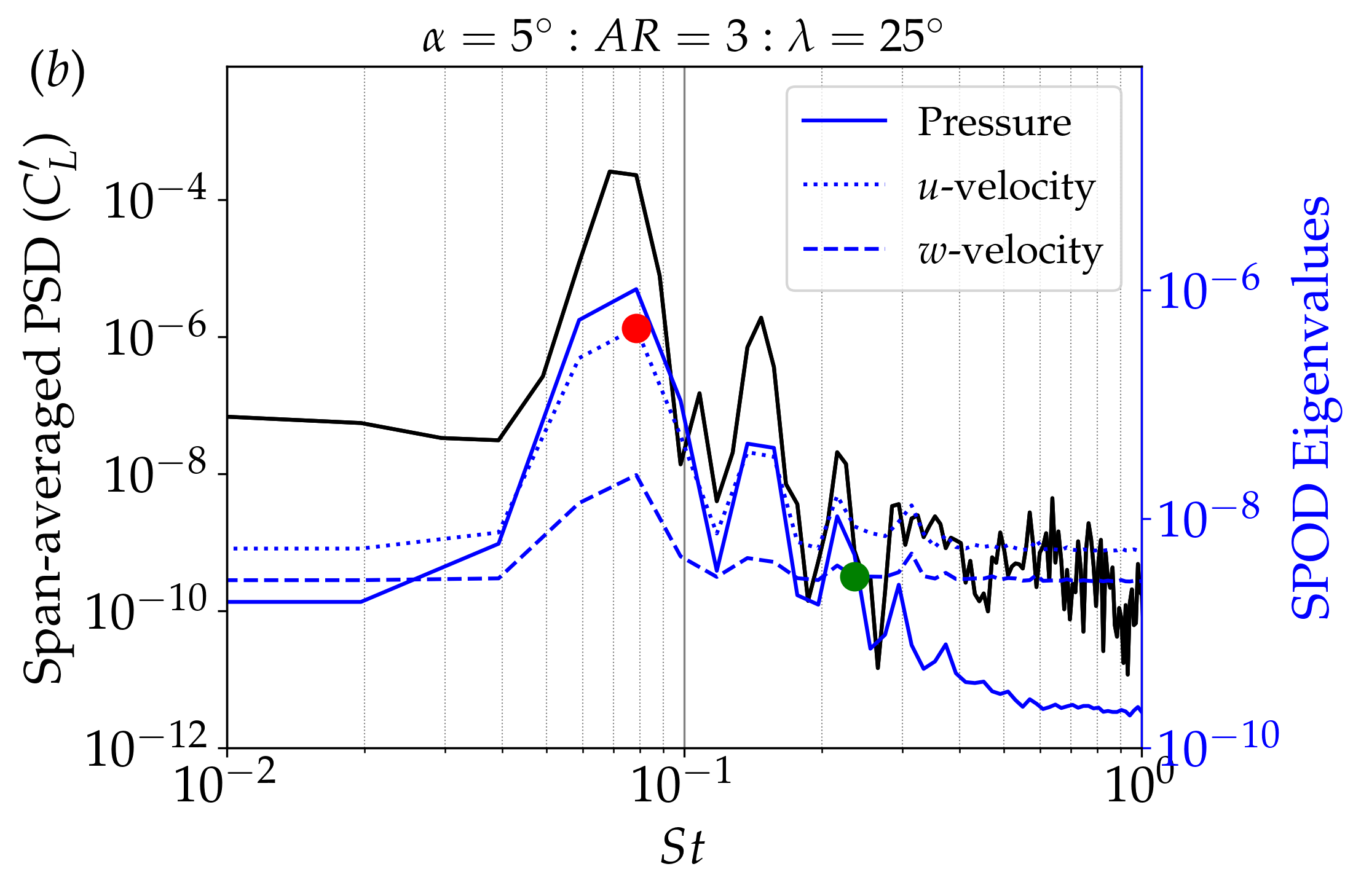}
\end{center}
\caption{SPOD spectra for pressure (blue, solid), $u$-velocity (blue, dotted) and $w$-velocity (blue, dashed), compared to the PSD evaluated on the fluctuations of the lift coefficient (black, solid). The peak frequencies corresponding to the 2D and 3D modes are indicated by red and green full circles, respectively. Showing sweep angles $\lambda = \left[0^{\circ}, 35^{\circ}\right]$ at an angle of attack of $\alpha = 5^{\circ}$.}
\label{fig:SPOD_spectra_AoA5}
\end{figure}

We first turn our attention to modal analysis of the quasi-2D case at $\alpha=5^{\circ}$ and $AR=3$ from Section~\ref{subsec:AoA5} that shows only intermittent three-dimensionality, primarily near the trailing edge (Fig.~\ref{fig:AoA5_surface_contours}). SPOD eigenvalue spectra are plotted in Fig.~\ref{fig:SPOD_spectra_AoA5} for $(a)$ $\lambda=0^{\circ}$ and $(b)$ $\lambda=25^{\circ}$ over three different flow variables. The pressure and $u$-velocity spectra correctly identify the low-frequency shock mode at $St=0.08$ in excellent agreement with the spectra obtained from $C^{\prime}_{L}$ for both cases. The spanwise $w$-velocity SPOD eigenspectra is completely flat at $\lambda=0^{\circ}$, reaffirming the quasi two-dimensionality of this lower angle of attack. The only other prominent peaks in the two spectra ($St = 0.16, 0.24$) occur at higher harmonics (not shown) of the 2D shock mode ($St=0.08$).

\begin{figure}
\begin{center}
\includegraphics[width=0.497\columnwidth]{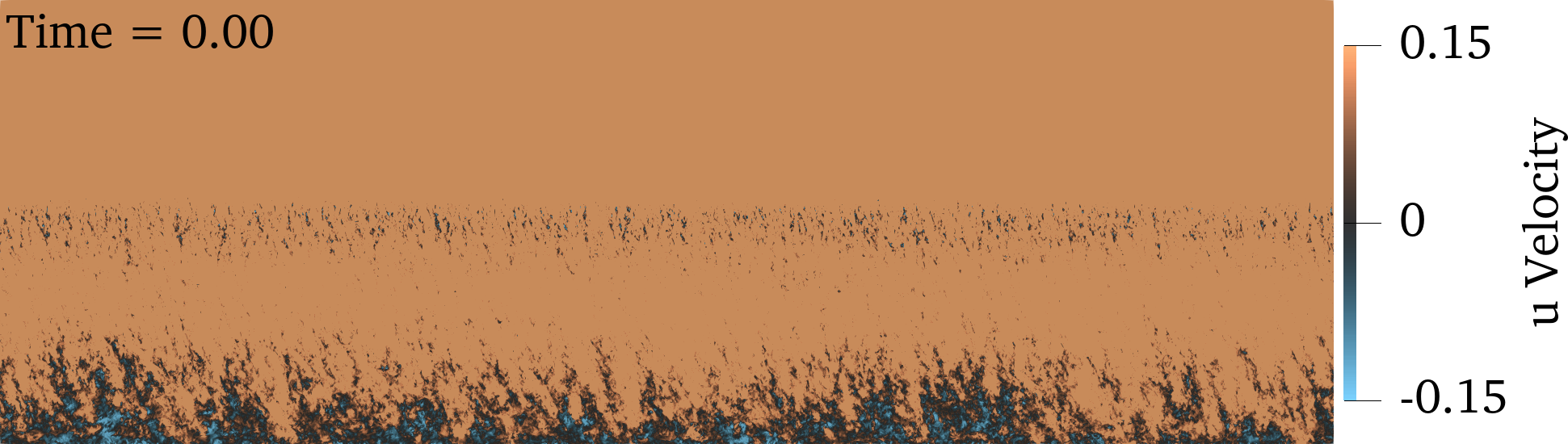}
\includegraphics[width=0.497\columnwidth]{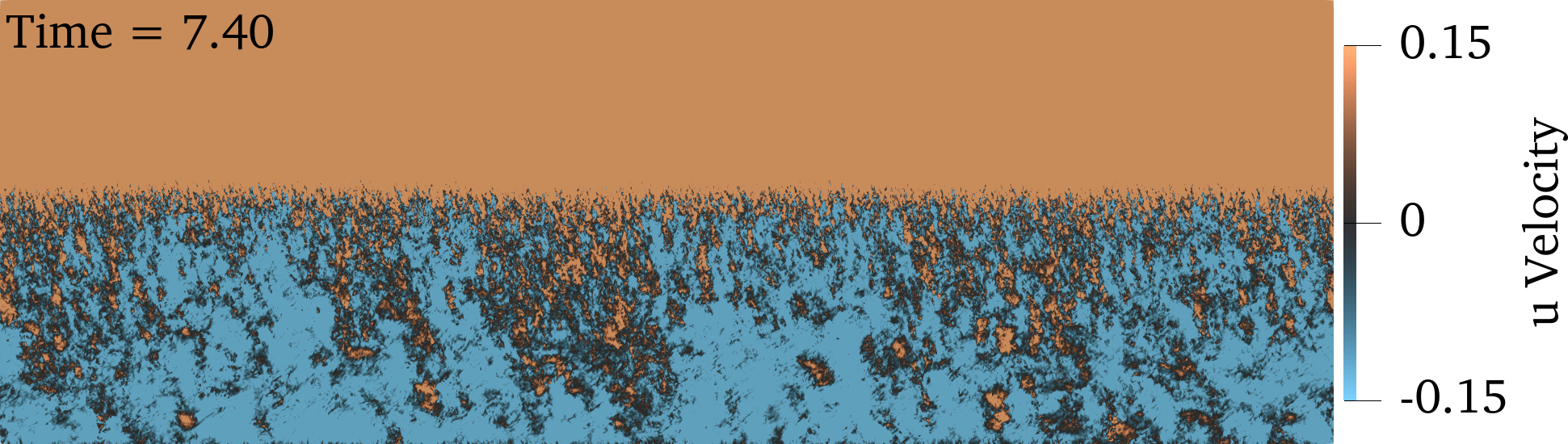}

\includegraphics[width=0.497\columnwidth]{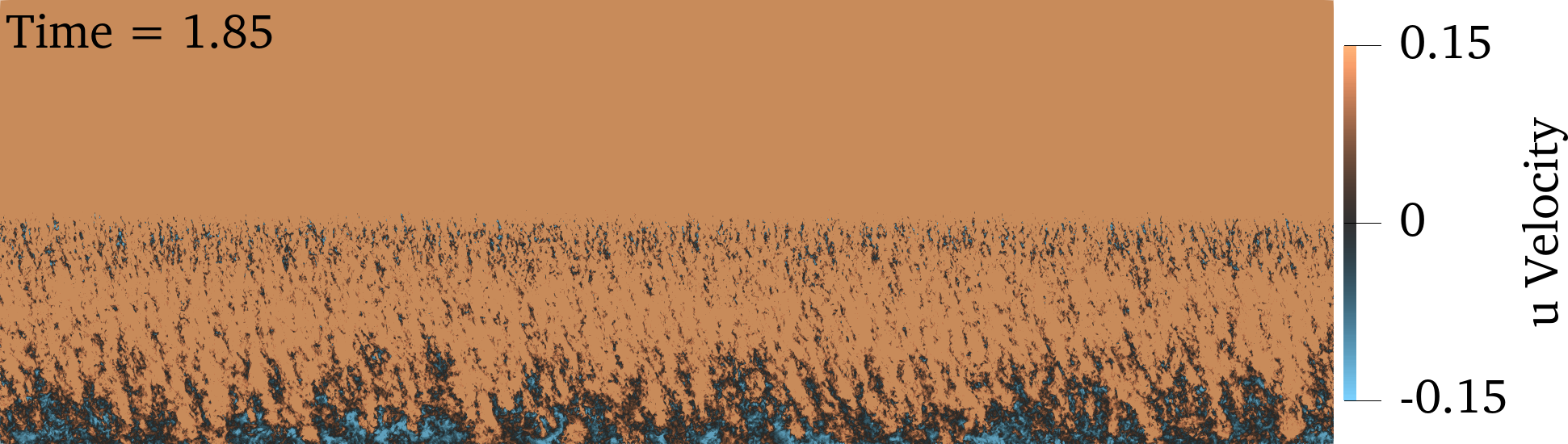}
\includegraphics[width=0.497\columnwidth]{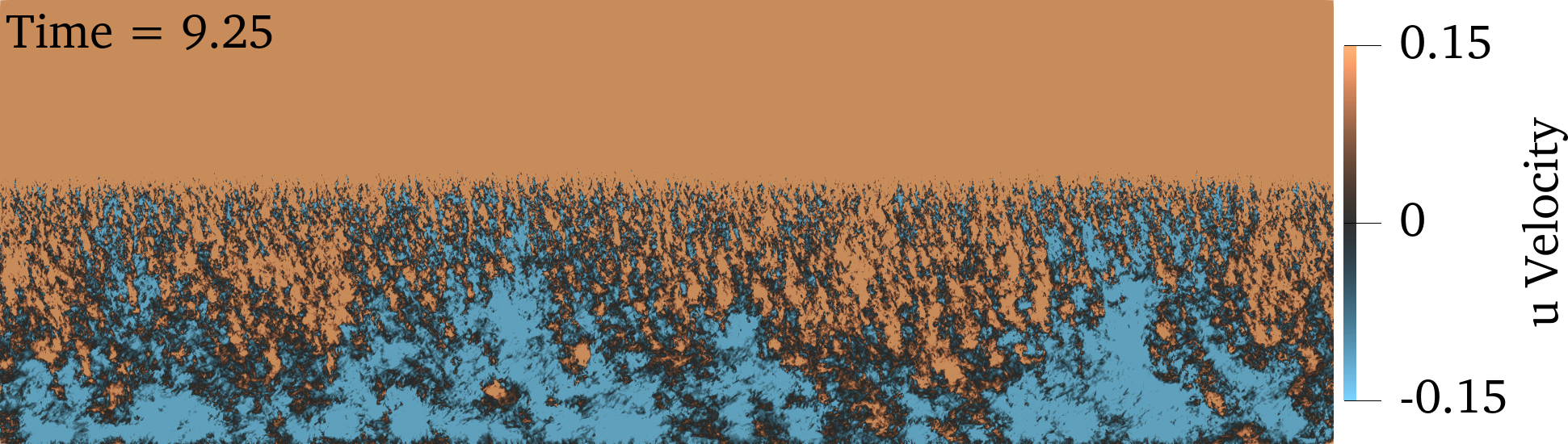}

\includegraphics[width=0.497\columnwidth]{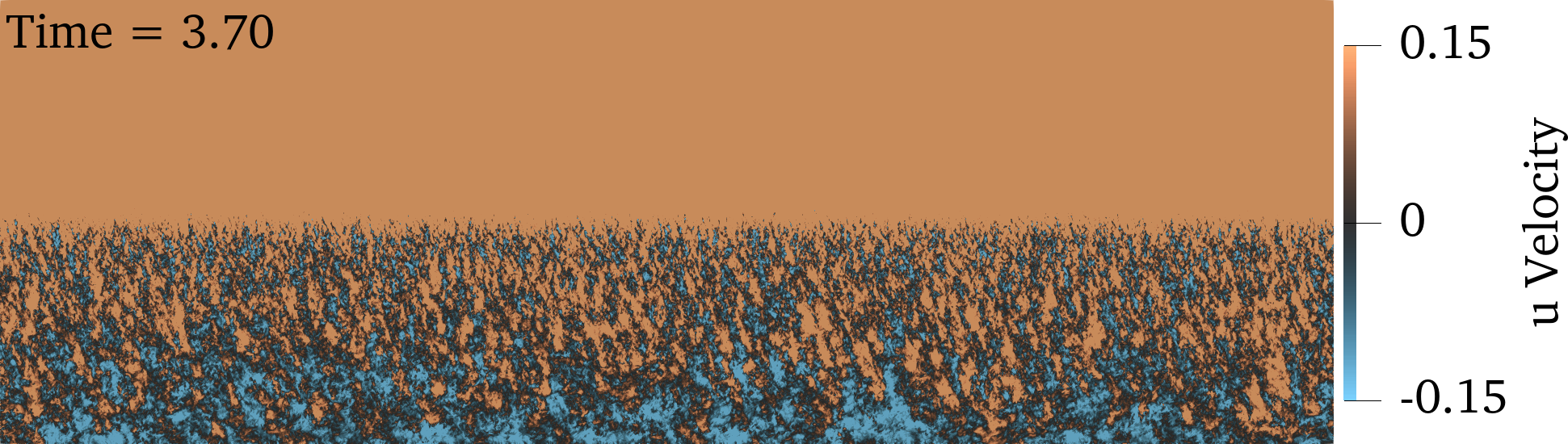}
\includegraphics[width=0.497\columnwidth]{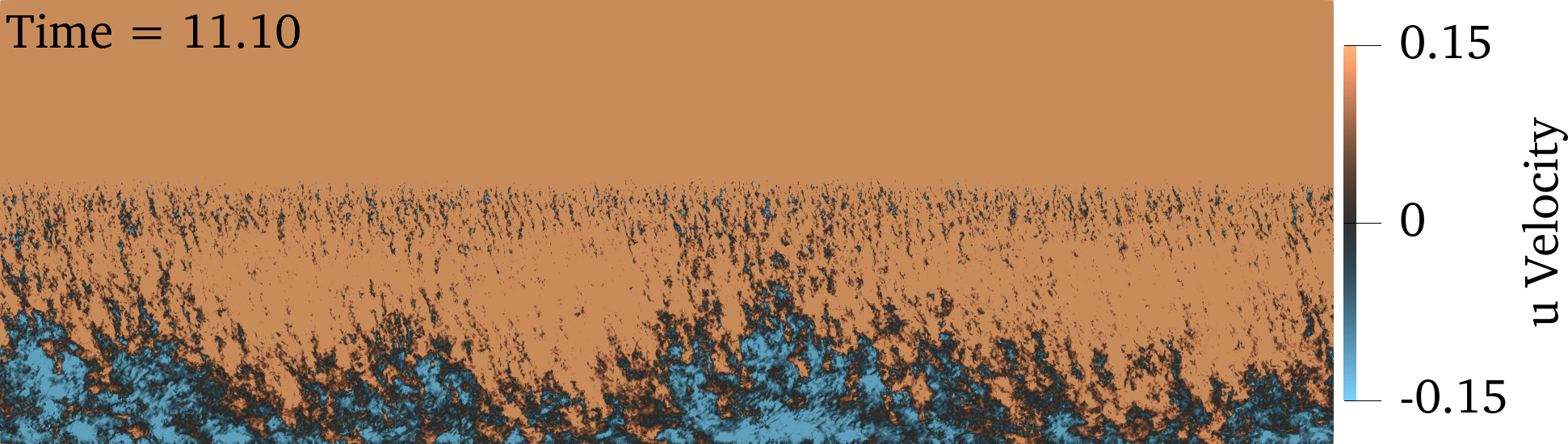}

\includegraphics[width=0.497\columnwidth]{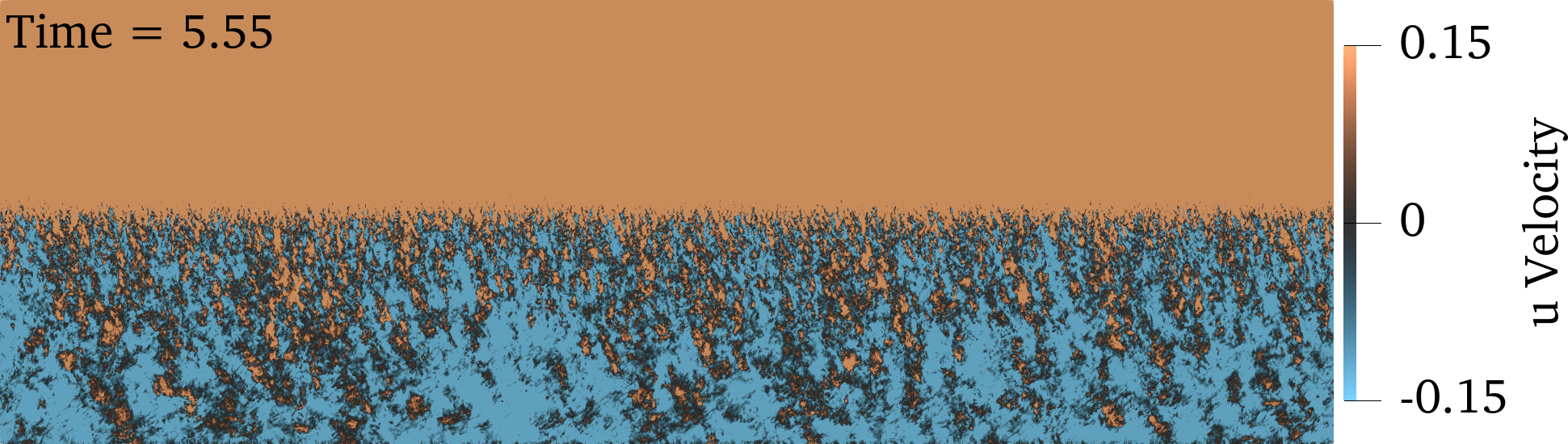}
\includegraphics[width=0.497\columnwidth]{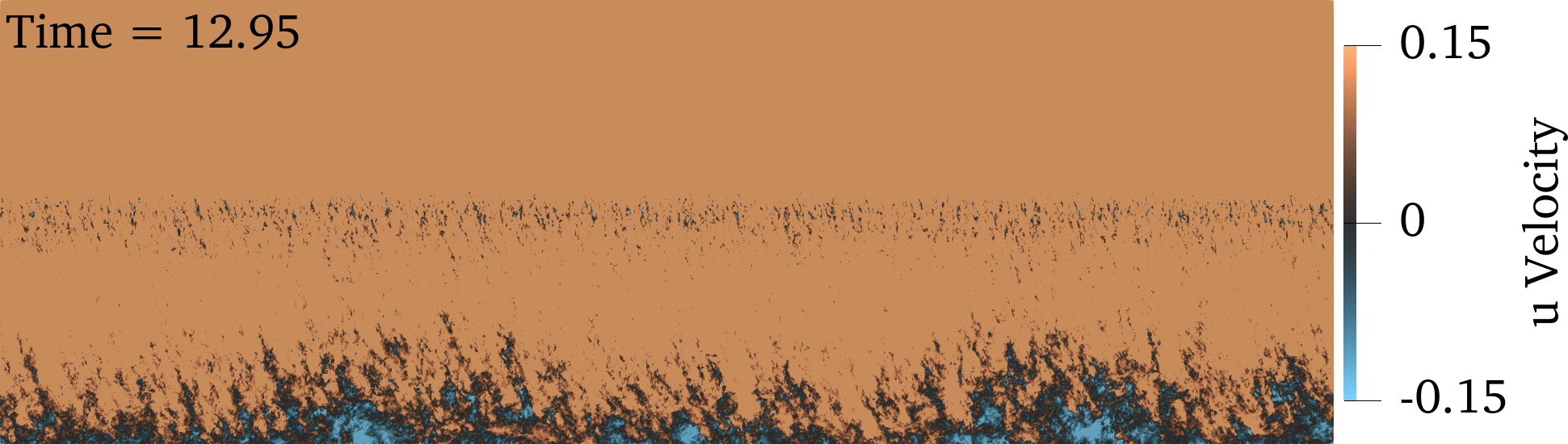}
\end{center}
\caption{Instantaneous streamwise velocity contours for the $\alpha = 5^{\circ}$ case at $AR=3$ and $\lambda=25^{\circ}$ showing the extent of the instantaneous reverse flow (blue) over one period of the $St \sim 0.08$ low-frequency buffet cycle t.}
\label{fig:AoA5_period}
\end{figure}

To investigate the presence of weak three-dimensionality in this case, Fig.~\ref{fig:AoA5_period} first shows instantaneous streamwise velocity contours from the $\alpha=5^{\circ}$ time series at $\lambda=25^{\circ}$ sweep. The view is the same as that in Fig.~\ref{fig:AoA5_surface_contours}, but here the contour range is restricted to $u / U_\infty \pm 0.15$ to highlight the spatial distribution of local chordwise flow reversal and better comment on the existence of spanwise three-dimensionality within the boundary layer. Eight snapshots are displayed over one full period of the low-frequency buffet cycle ($St = 0.077$). At $t=t_0$, the majority of the flow over the airfoil is attached at the high-lift phase of the cycle. At mid-chord, only small-scale turbulence fluctuations are observed at $u/U_\infty \approx 0$ along the shock position. The trailing edge is more strongly separated, with small pockets of local chordwise flow reversal present in light blue. The skewed orientation of the flow structures relative to the direction of the chord is apparent, as a result of the imposed sweep angle. In-between the shock position and trailing edge, a wide strip of fully-attached flow is present (as in the unswept $C_f$ distribution in Fig.~\ref{fig:deg5_lines}$(b)$ at $x \approx 0.65$).

Over the next quarter of the cycle ($t \sim 3.7$), the spatial extent of the flow separation increases greatly to cover the entire region downstream of the SBLI. The strongest flow reversal occurs around the trailing edge, with fluid moving upstream towards the shock. By mid-cycle ($t=5.5-7.4$), the back half of the airfoil is fully-separated. The extensive flow separation covers the entire region from the back of the shock to the trailing edge. Despite the span width ($L_z = 3c$) being wide enough to accommodate the buffet cell instability ($\lambda_z = 1-1.5c$) and the presence of significant flow separation, the shock front is still essentially-2D across the span, normal to the leading edge. Small protrusions of the shock front are visible upon close inspection at $t=7.4$, but they are extremely minor. This is very different behaviour to the higher angle of attack case at $\alpha=6^\circ$, in which strong spanwise-travelling perturbations of the shock front are observed (Fig.~\ref{fig:buffet_cells}) at all times in the time series.

The buffet mechanism observations from the simulations and resolvent analysis of \citet{KAWAI_RESOLVENT2023} between shock strength, position, and trailing edge separation height, indicate that at this time ($t=7.4$) the shock is farthest upstream when the flow separation reaches its maximum. \blue{The relative shock strength can be computed at different times from the pre- and post-shock pressure values $(p_1, p_2)$ with $\gamma=1.4$ as
\begin{equation}
M_r = \sqrt{\,1 + \frac{\gamma + 1}{2\gamma}\left(\frac{p_2}{p_1} - 1\right)\,}.
\end{equation}
During the low-lift phase, the relative shock strength decreases by 8\% compared to the point of maximum lift. The weakening of the shock strength results in a reduction in the adverse pressure gradient felt by the boundary layer}, causing the flow to start reattachment. As flow separation begins to subside ($t=9.25$ in Fig.~\ref{fig:AoA5_period}), traces of three-dimensionality are visible in the remnants of the separated flow contours at the same $\lambda_z = 1.5c$ spanwise wavelength of the buffet cells at $\alpha=6^{\circ}$. These 3D structures convect downstream with the freestream ($t=11.10$), before being shed downstream ($t=12.95$) in the wake of the airfoil. Here, the cycle is complete, and the flow is back to its original, mostly attached state (as in $t=0$).

The flow visualizations show that, even in the quasi-2D buffet case, weak spanwise perturbations do exist in the separated flow region during certain points of the low-frequency buffet cycle. However, they \red{do not cause the appearance of buffet cells nor the spanwise modulation of} the shock wave, as was the case for the more heavily separated $\alpha=6^{\circ}$ angle of attack configuration. When the time/span-averaged flow separation is minimal, the chordwise shock oscillations (2D mode, $St\sim0.08$) dominate, and the flow behaves in a quasi-2D manner, even when the airfoil aspect ratio is large enough to accommodate the 3D separation cell mode. The 3D mode is therefore extremely weak at $\alpha=5^{\circ}$, and difficult to identify even with long time series and high-fidelity simulation. The relative energy content of the 3D features are at least three orders of magnitude lower than the 2D shock mode (Fig.~\ref{fig:SPOD_spectra_AoA5}$(b)$) in this case.

While quasi-2D buffet dynamics can exist on wide-span infinite wings in certain circumstances when the flow separation is weak, the flow becomes fully-3D once the angle of attack is raised. The 3D aspect of the buffet instability appears to be primarily due to the existence and topology of the local flow separation, in agreement with our unswept buffet study \citep{lusher2025highfidelity_JFM}, and recent modal analysis work applied to experimental databases of full aircraft configurations by \citet{VMD_Ohmichi2024}. The shock does, however, play an important role in causing the boundary layer to separate in the first place at moderate AoA, generating the necessary conditions for prominent 3D buffet cells to appear. Once there is sufficient flow separation present, the energy of the 3D mode is highly localised to the region containing the shock and its associated chordwise motion, but this is not the case for the $\alpha=5^{\circ}$ case here.

\begin{figure}
\begin{center}
\includegraphics[width=0.497\columnwidth]{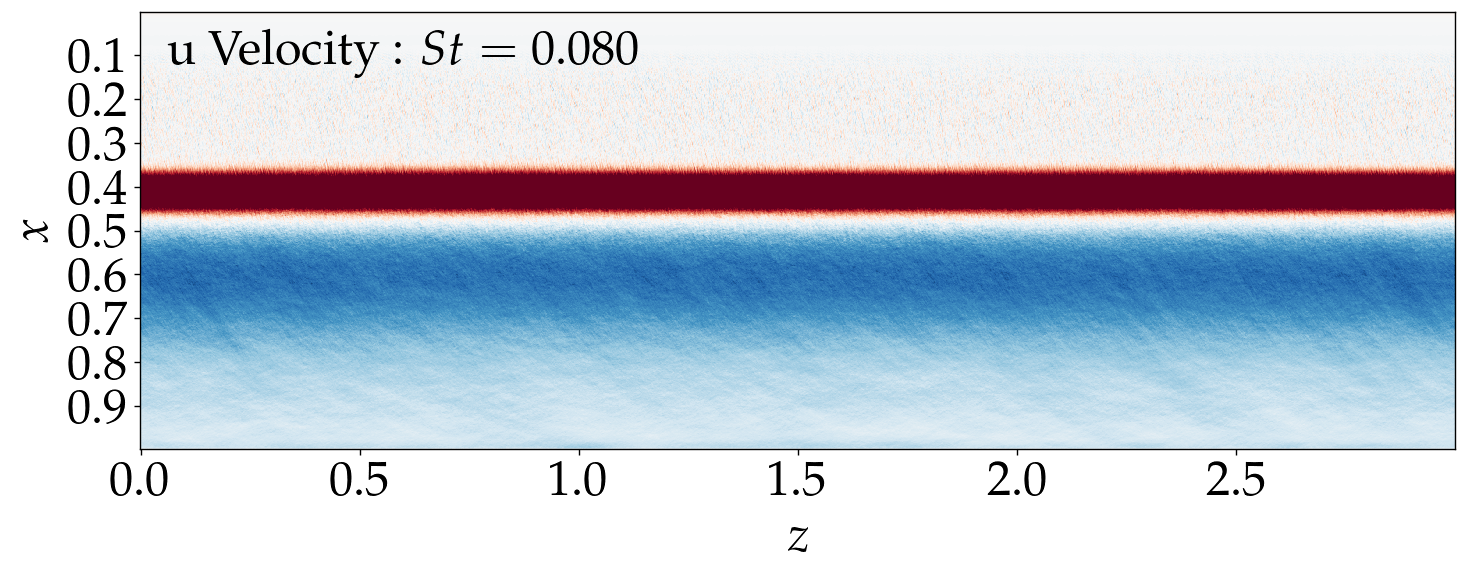}
\includegraphics[width=0.497\columnwidth]{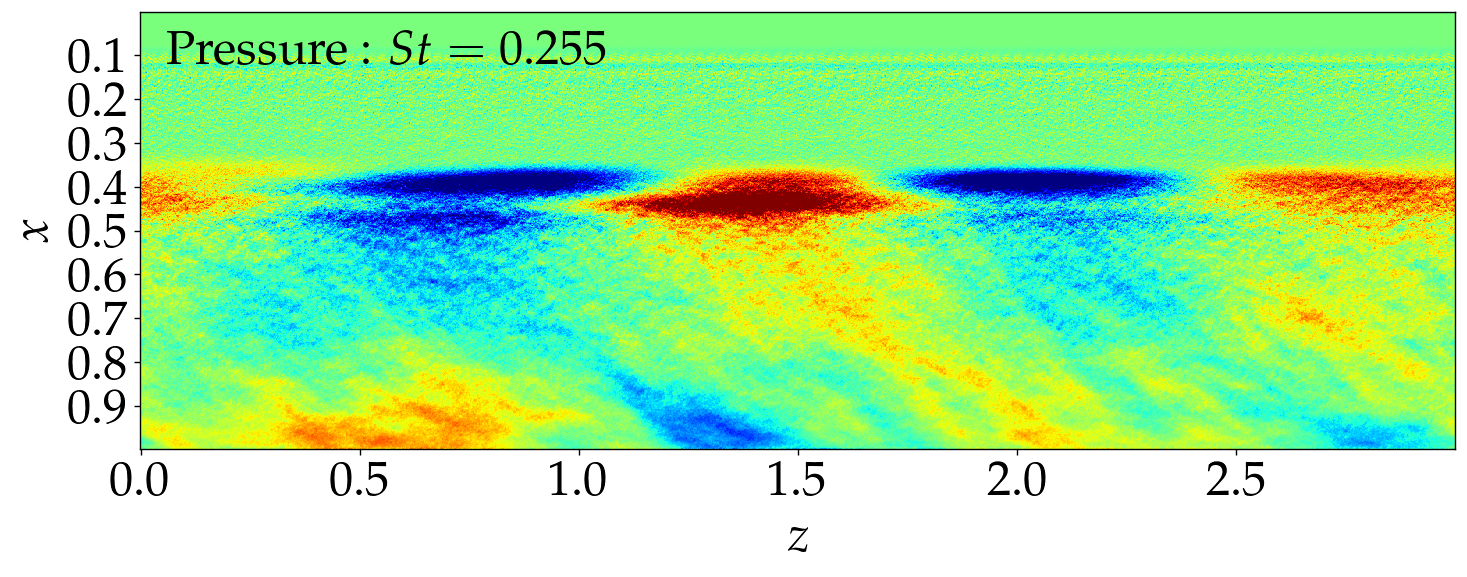}
\includegraphics[width=0.497\columnwidth]{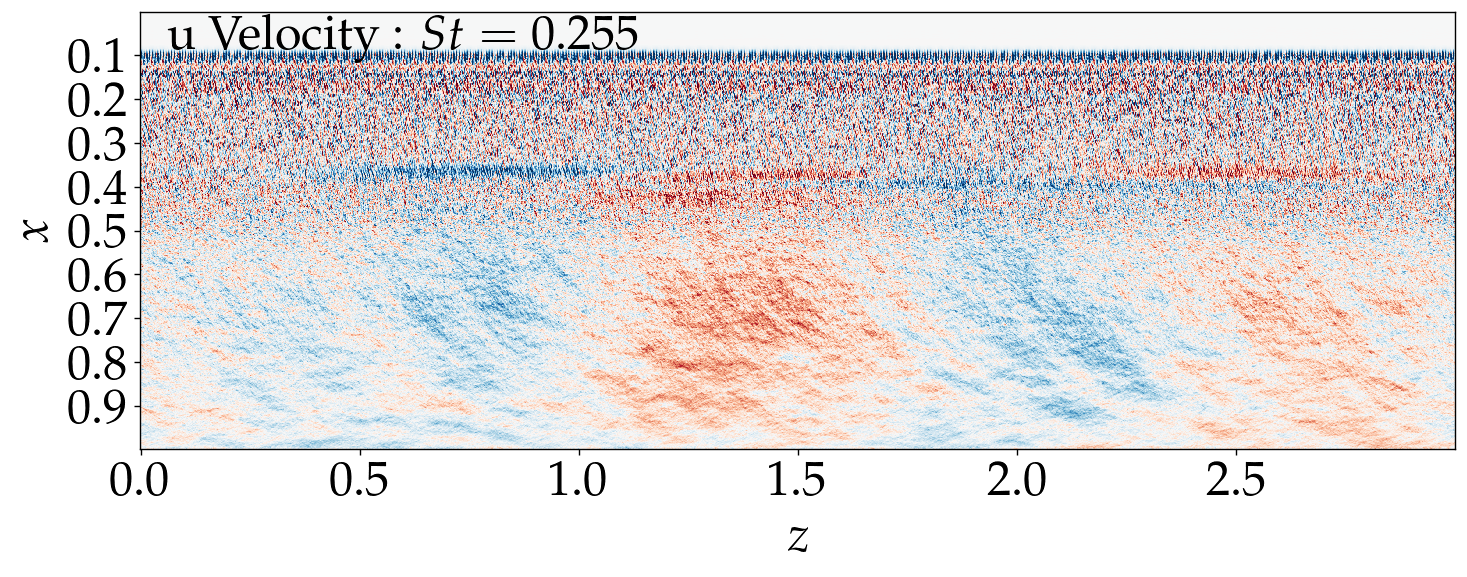}
\includegraphics[width=0.497\columnwidth]{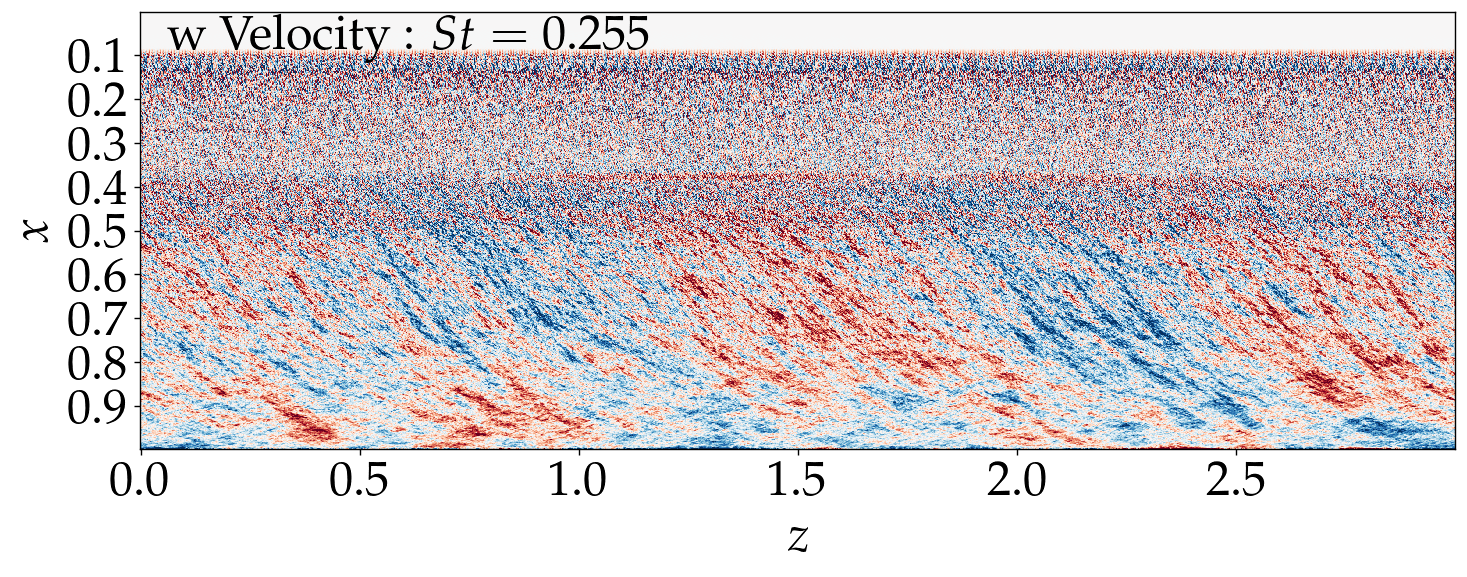}
\end{center}
\caption{SPOD modes for the $\alpha = 5^{\circ}$ case at $\lambda = 25^{\circ}$ sweep. Showing the ($St = 0.08$) 2D chordwise shock oscillation mode and \orange{($St=0.255$)} 3D spanwise-convecting buffet cell mode, based on pressure, \red{$u$-velocity, and $w$-velocity} flow field data.}
\label{fig:SPOD_modes_contours_AoA5}
\end{figure}

The swept $\lambda=25^{\circ}$ case at $\alpha=5^{\circ}$ was run twice as long $(t - t_0 \sim 100)$ as the unswept $\lambda=0^{\circ}$ one in an attempt to identify the faint intermittent three-dimensionality detected in the instantaneous flow fields near the trailing edge. However, we see that the effect on the $C_{L}^{\prime}$ and SPOD eigenvalue spectra (Fig.~\ref{fig:SPOD_spectra_AoA5}$(b)$) is mainly improved resolution of the shock mode and its higher harmonics. The $w$-velocity spectra also exhibits minor peaks around the shock mode and the higher harmonics ($St = 0.16, 0.24, 0.32$). This suggests that, while there is no obvious buffet cell mode as was reported in the $\alpha=6^{\circ}$ case at the same sweep angle (Fig.~\ref{fig:SPOD_spectra}($(e)$, detected in all three variables), the $w$-velocity SPOD does show very minor latent spanwise energy coupled to the chordwise motion of the shock.

Finally, Fig.~\ref{fig:SPOD_modes_contours_AoA5} displays the SPOD modes for this case at $St=0.08$ and $St=0.255$. The $St=0.255$ value was selected as the closest available mode to match the 3D peak in the $C_{L}^{\prime}$ spectra (Fig.~\ref{fig:deg6_lines_low}$(d)$) with the same sweep strength at the higher AoA. Other intermediate modes were examined around this range ($0.2 < St < 0.4$) to find the most illustrative one. At $St=0.08$, the 2D shock mode is entirely independent of the span, with no trace of large-scale three-dimensionality. Although the $St=0.255$ mode does not show a notable peak in the spectra (Fig.~\ref{fig:SPOD_spectra_AoA5}$(b)$), the SPOD modes reveal a very weak spanwise perturbation of the expected $\lambda_z = 1.5c$ buffet cell wavelength (at $AR=3$). \red{The $u$-velocity decomposition exhibits a spanwise mode shape concentrated at both the shock ($x \sim 0.35$) and the region downstream of the SBLI (centred at $x \sim 0.7$)}. \blue{The two regions are seen to be in phase with each other for this variable along a given line in the chordwise direction}. \red{The $w$-velocity meanwhile, shows a similar oblique spanwise-repeating mode pattern downstream of the SBLI ($x \sim 0.7$), but no macroscopic structures at the shock itself.} The pressure field further highlights the alternating phase cellular mode shapes along the shock, which convect across the span (SPOD mode animations are provided in the supplementary material \cite{supplementary_material}).

This shows that, although the instantaneous flow acts in a quasi-2D manner at this lower angle of attack where the time-averaged flow is mostly attached, there is still a very weak 3D spanwise-convecting component active that the SPOD is able to detect. This could also be seen as an weakly globally unstable mode, explaining why the RANS-based GSA studies found simultaneous onset of both 2D and 3D instabilities. Only when the angle of attack is increased and the flow separation increases, does the separation mode get amplified strongly enough in the shocked region, to lead to the strong three-dimensionality and concentration of energy around the intermediate frequencies as is reported for the appearance of buffet cells. \red{The spectral content of this scenario is examined in the next section.}

\subsection{Modal analysis of the 3D $(\alpha = 6^{\circ})$ buffet interaction}\label{subsec:AoA-6_SPOD}

\begin{figure}
\begin{center}
\includegraphics[width=0.495\columnwidth]{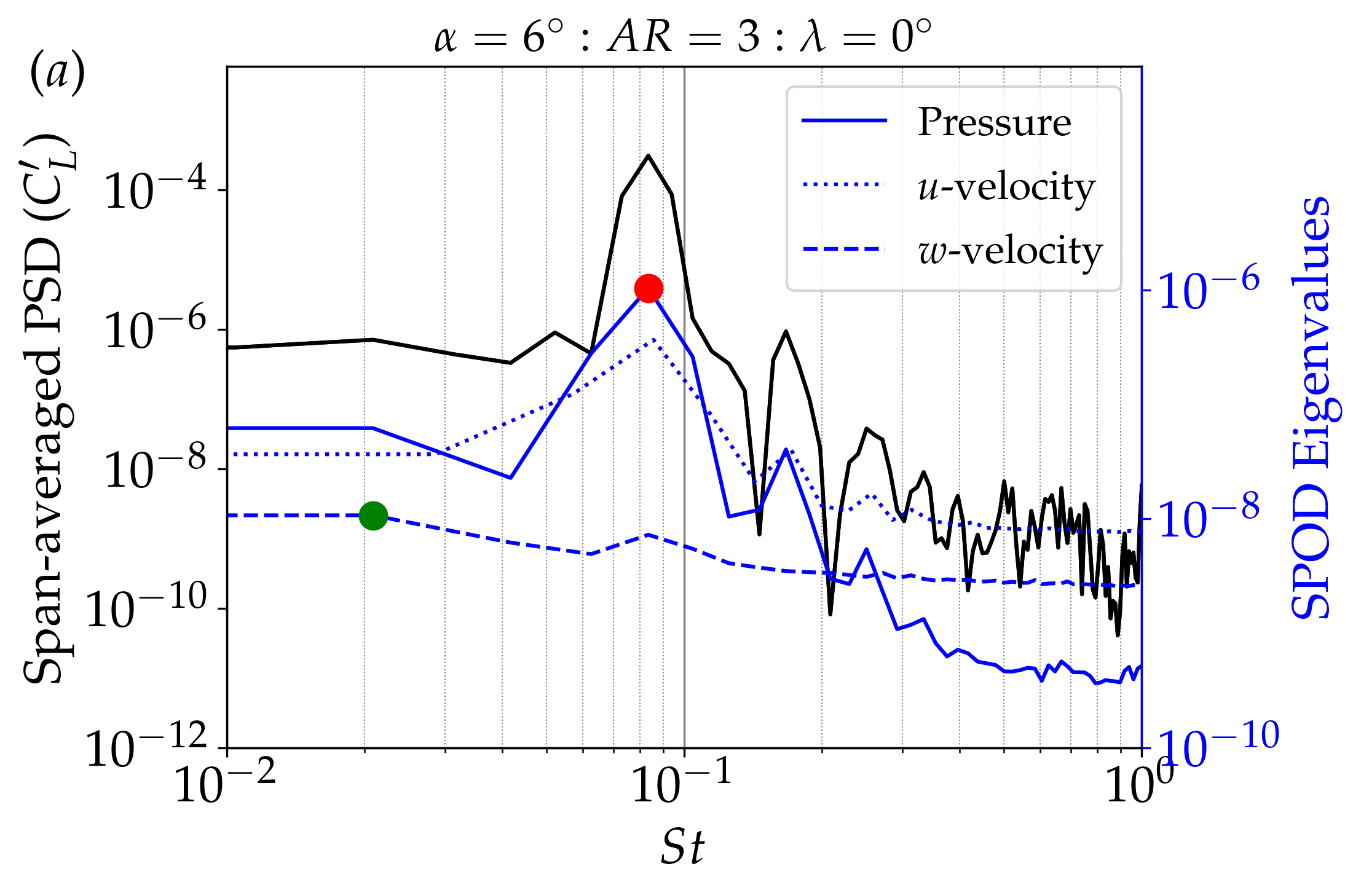}
\includegraphics[width=0.495\columnwidth]{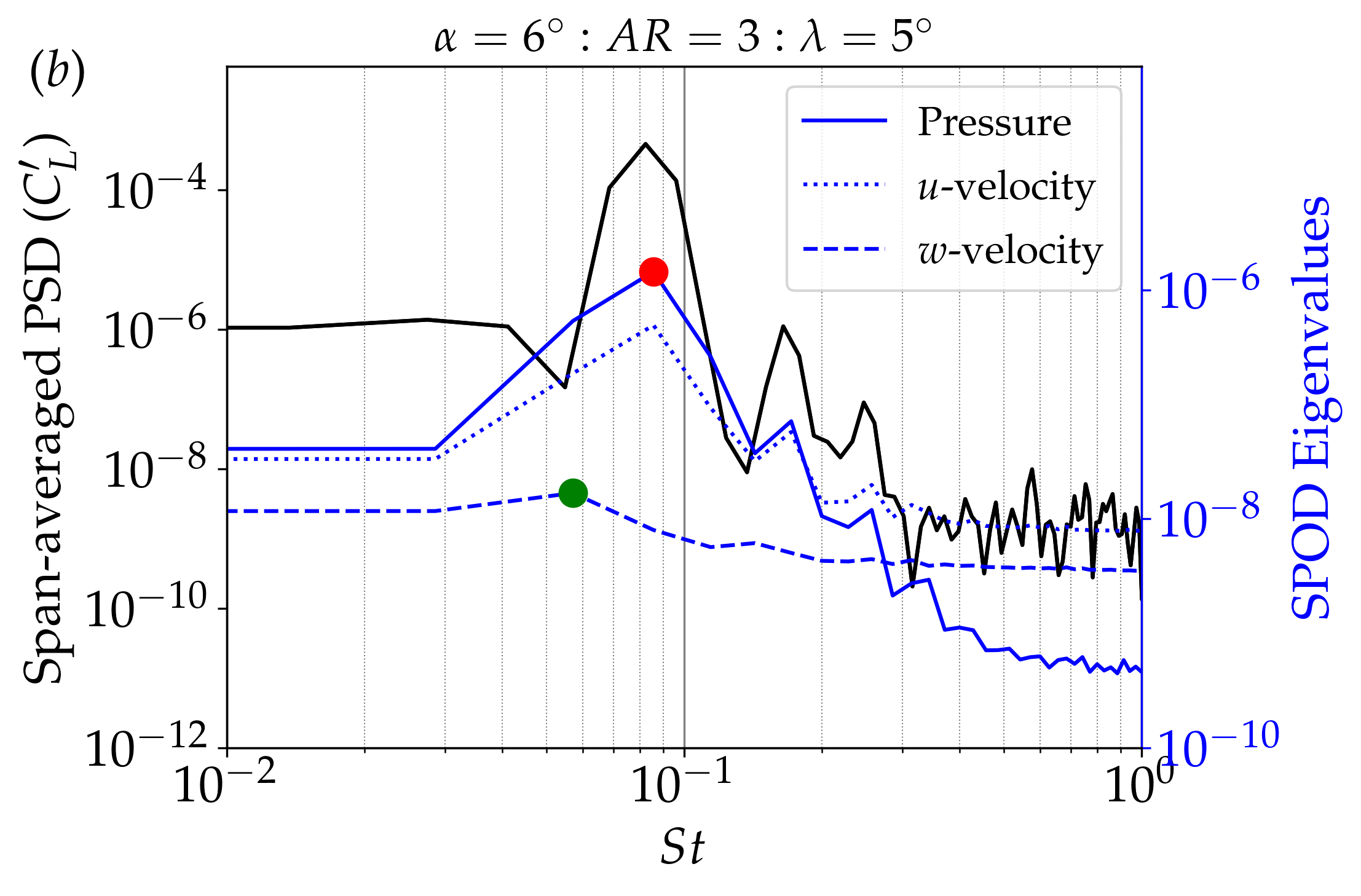}
\includegraphics[width=0.495\columnwidth]{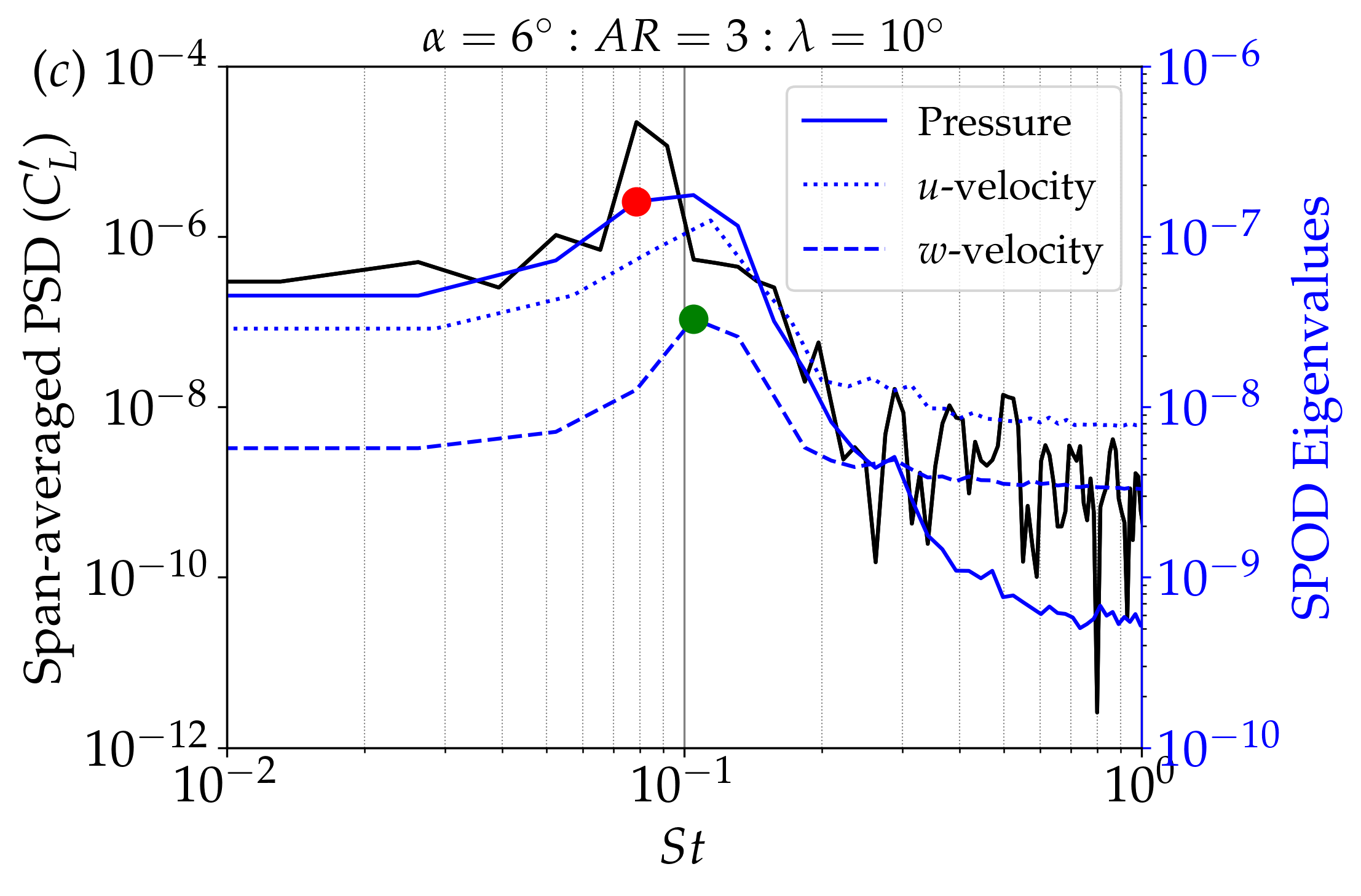}
\includegraphics[width=0.495\columnwidth]{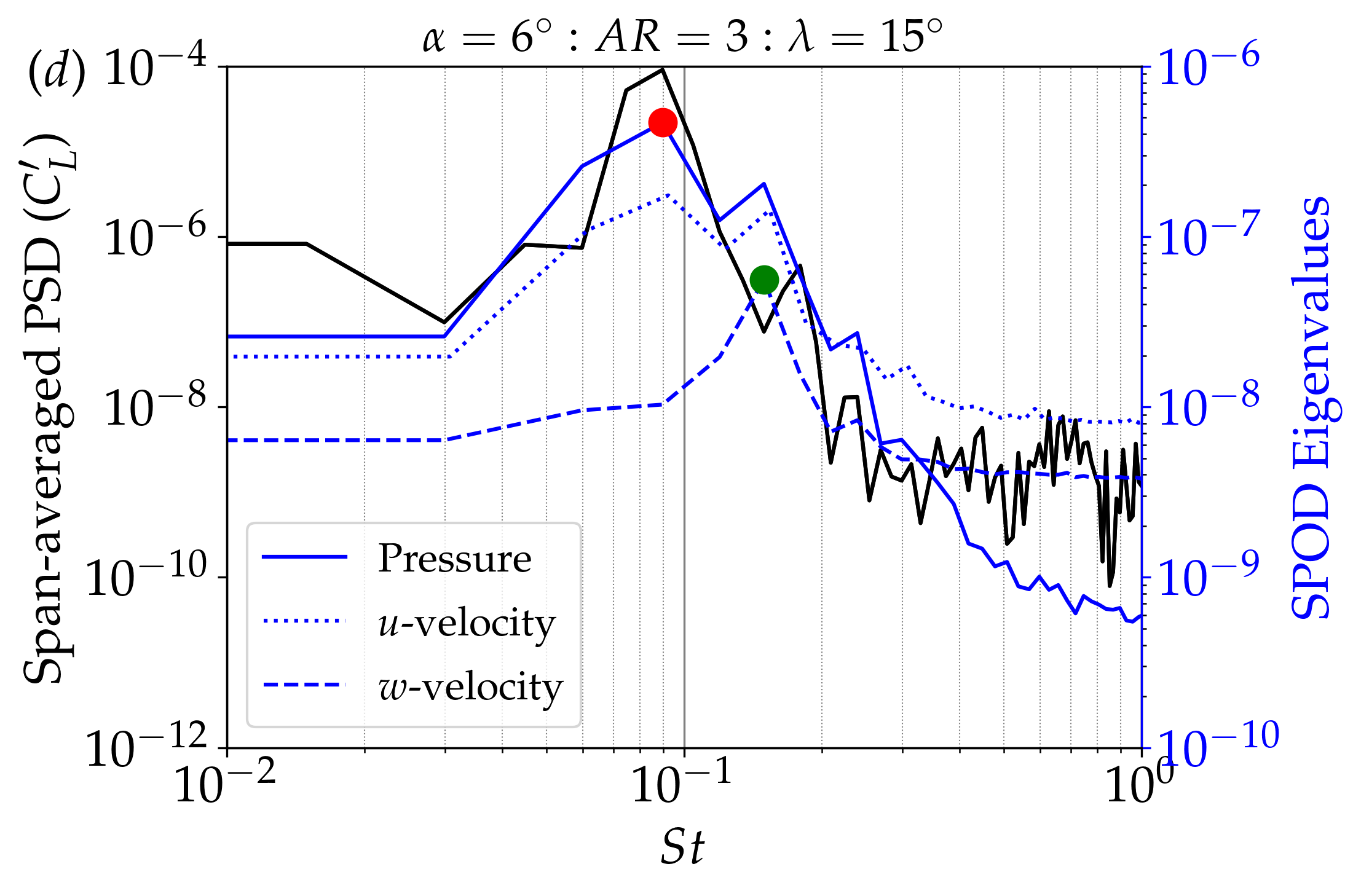}
\includegraphics[width=0.495\columnwidth]{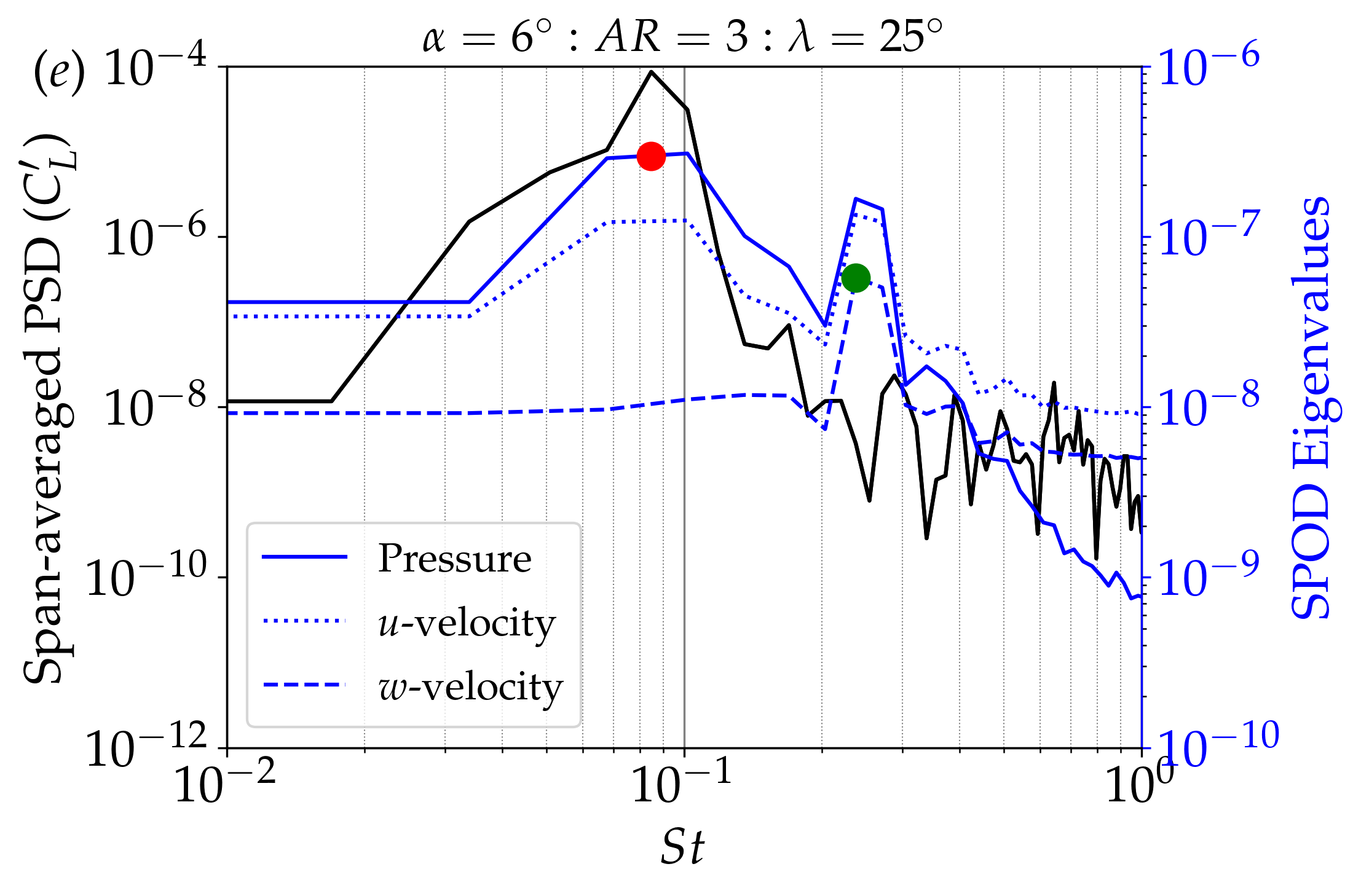}
\includegraphics[width=0.495\columnwidth]{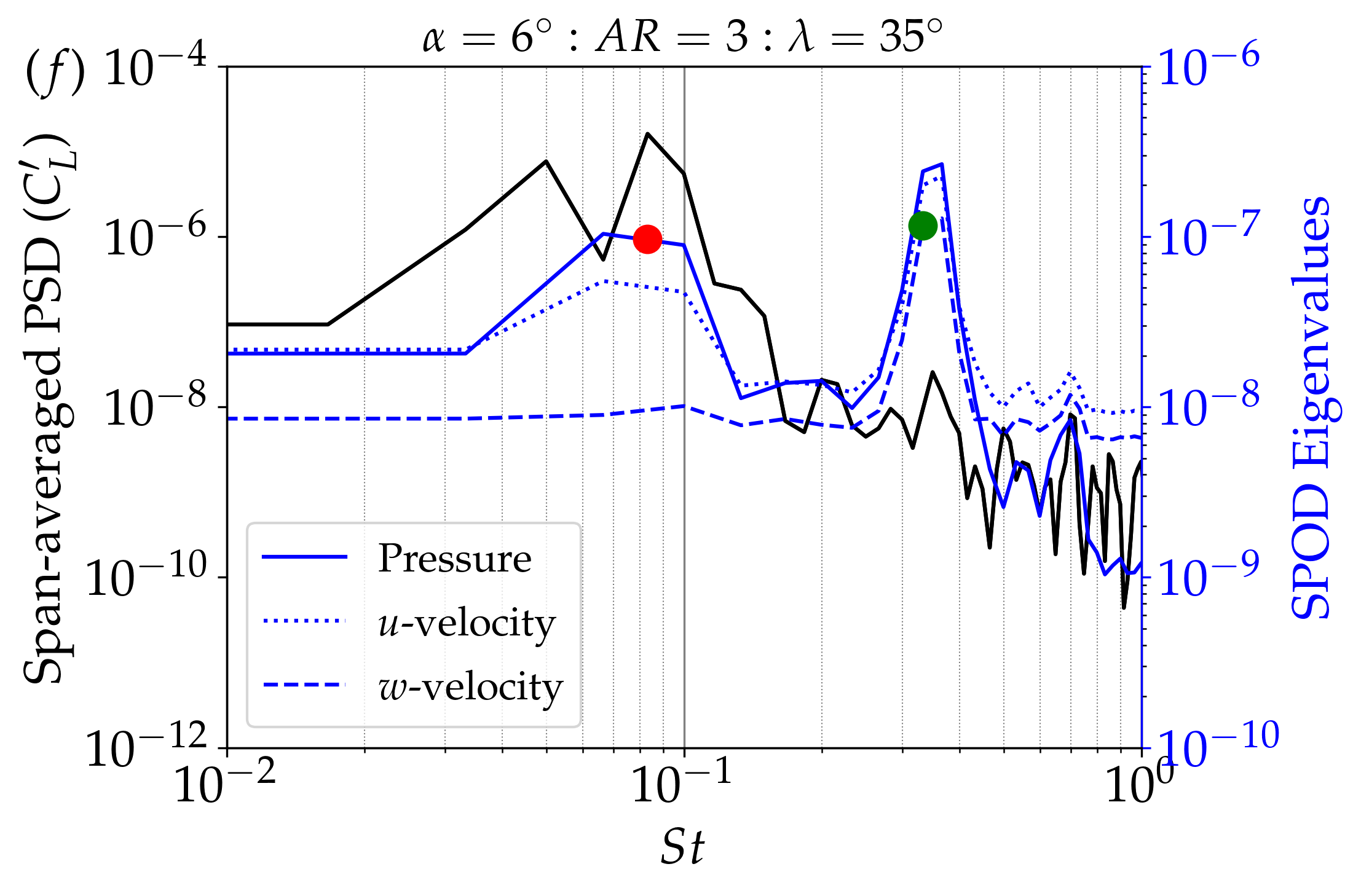}
\end{center}
\caption{SPOD spectra for pressure (blue, solid), $u$-velocity (blue, dotted) and $w$-velocity (blue, dashed), compared to the PSD of lift coefficient fluctuations (black, solid). The peak frequencies corresponding to the 2D and 3D modes are indicated by red and green full circles, respectively. Showing a range of sweep angles between $\lambda = \left[0^{\circ}, 35^{\circ}\right]$ at an angle of attack of $\alpha = 6^{\circ}$.}
\label{fig:SPOD_spectra}
\end{figure}

Figures~\ref{fig:SPOD_spectra}$(a)$-$(f)$ show the SPOD eigenvalue spectra at $\alpha=6^{\circ}$ for SPOD applied to the wall-pressure field (blue, solid), streamwise velocity, $u$ (blue, dotted), and spanwise velocity, $w$ (blue, dashed), compared to the PSD of span-averaged lift fluctuations (black, solid). All six sweep angles are considered between $\lambda = \left[0^{\circ}, 35^{\circ}\right]$. 

\begin{figure}
\begin{center}
\includegraphics[width=0.497\columnwidth]{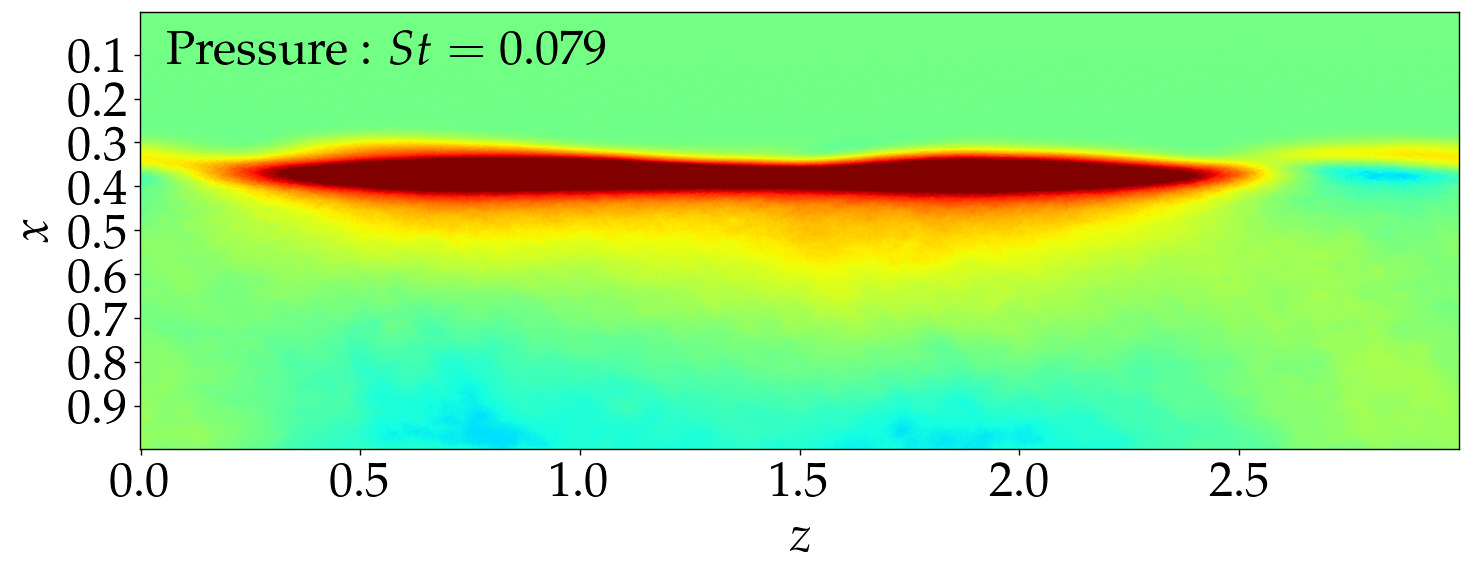}
\includegraphics[width=0.497\columnwidth]{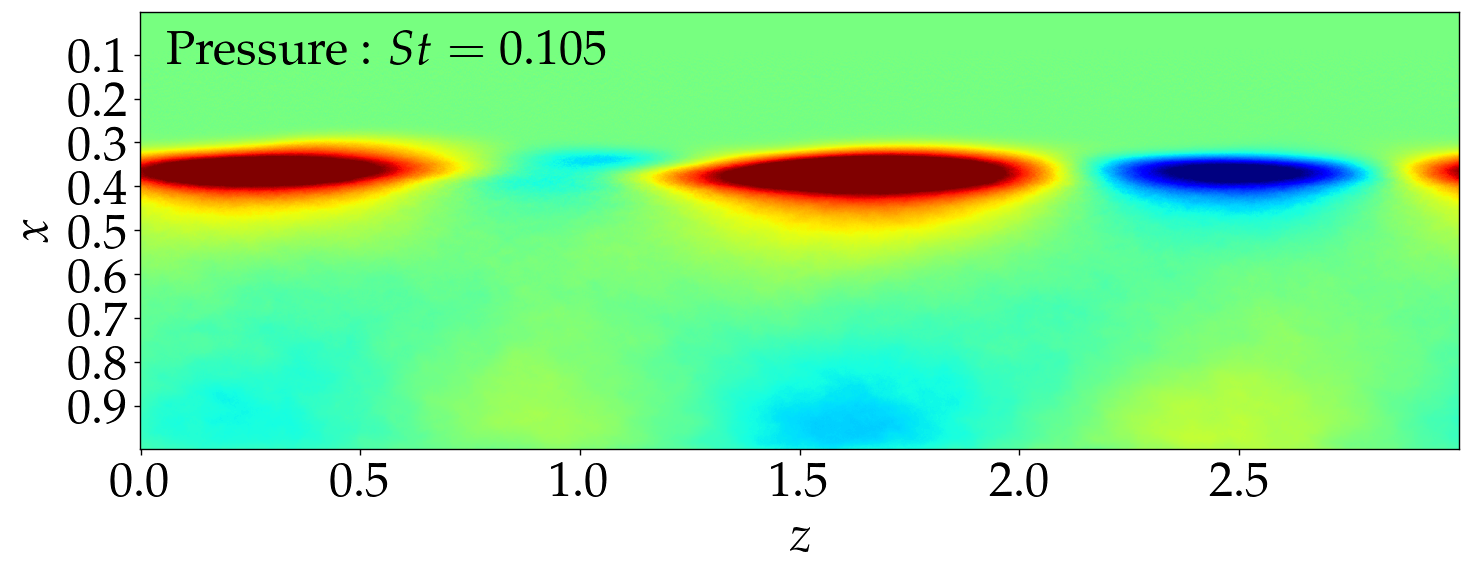}
\includegraphics[width=0.497\columnwidth]{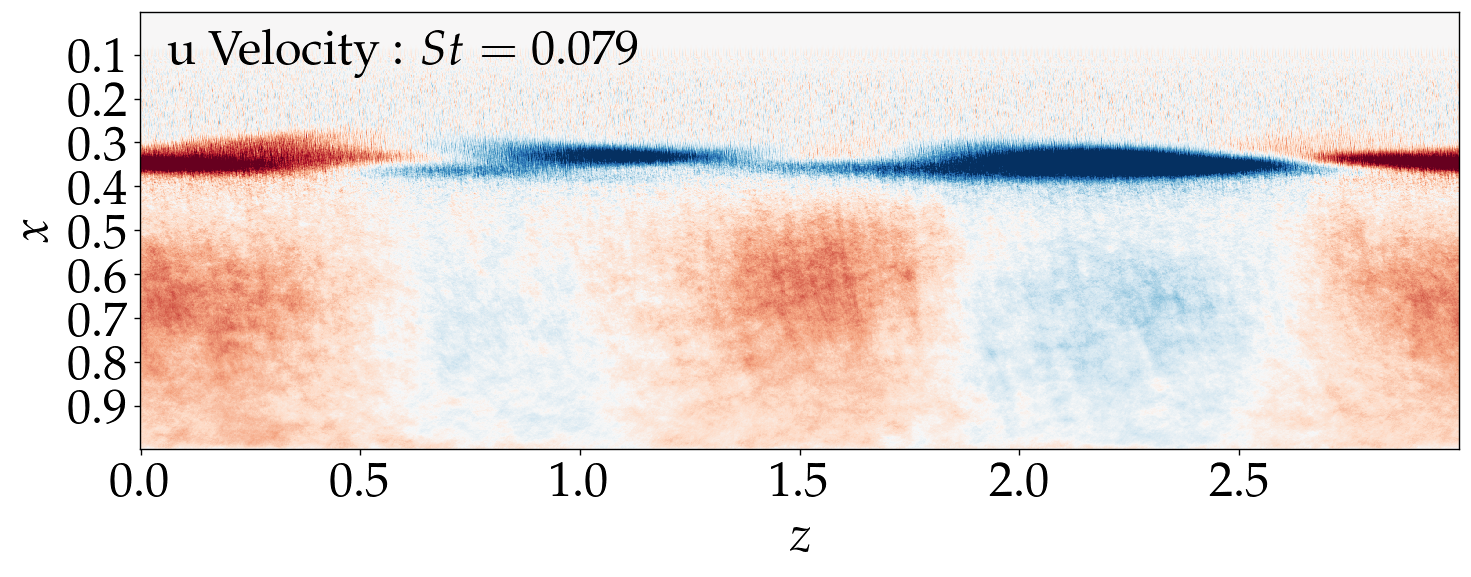}
\includegraphics[width=0.497\columnwidth]{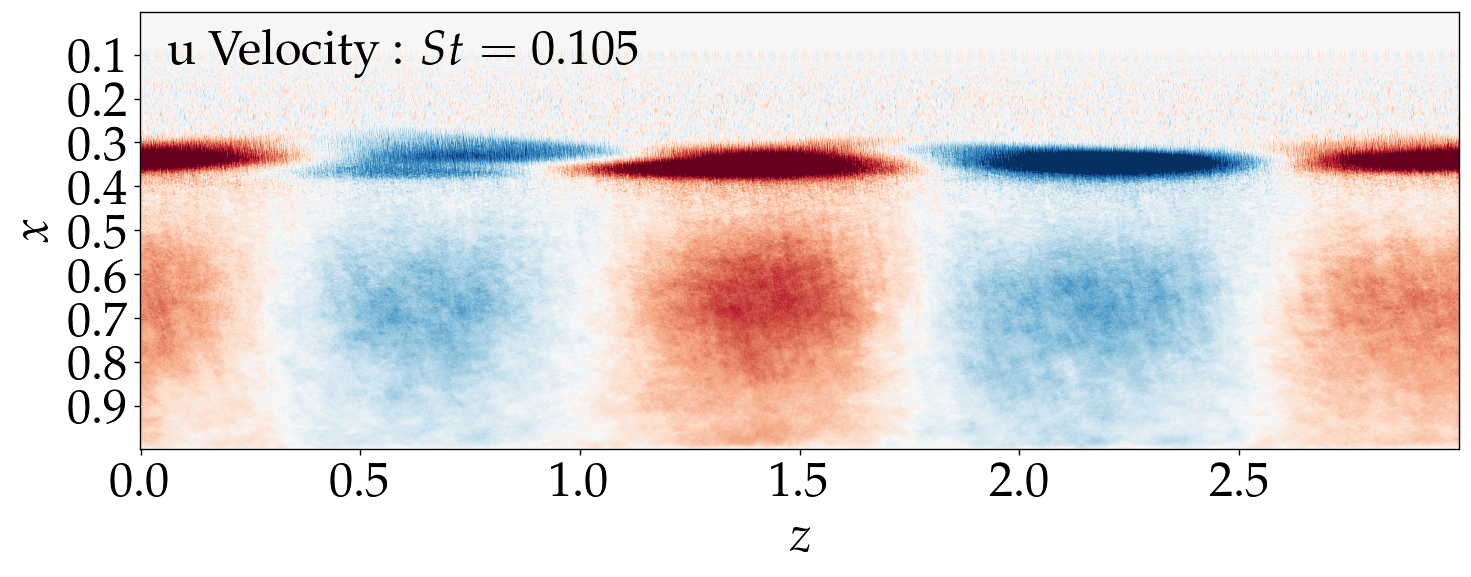}
\end{center}
\caption{\red{SPOD modes at $\alpha = 6^{\circ}$ $AR=3$ and $\lambda = 10^{\circ}$ (as in Fig.~\ref{fig:SPOD_spectra}$(c)$ spectra), for (top) pressure and (bottom) $u$-velocity, showing two different frequencies.}}
\label{fig:SPOD_modes_LAMBDA10_OVERLAP}
\end{figure}

\begin{figure}
\begin{center}
\includegraphics[width=0.497\columnwidth]{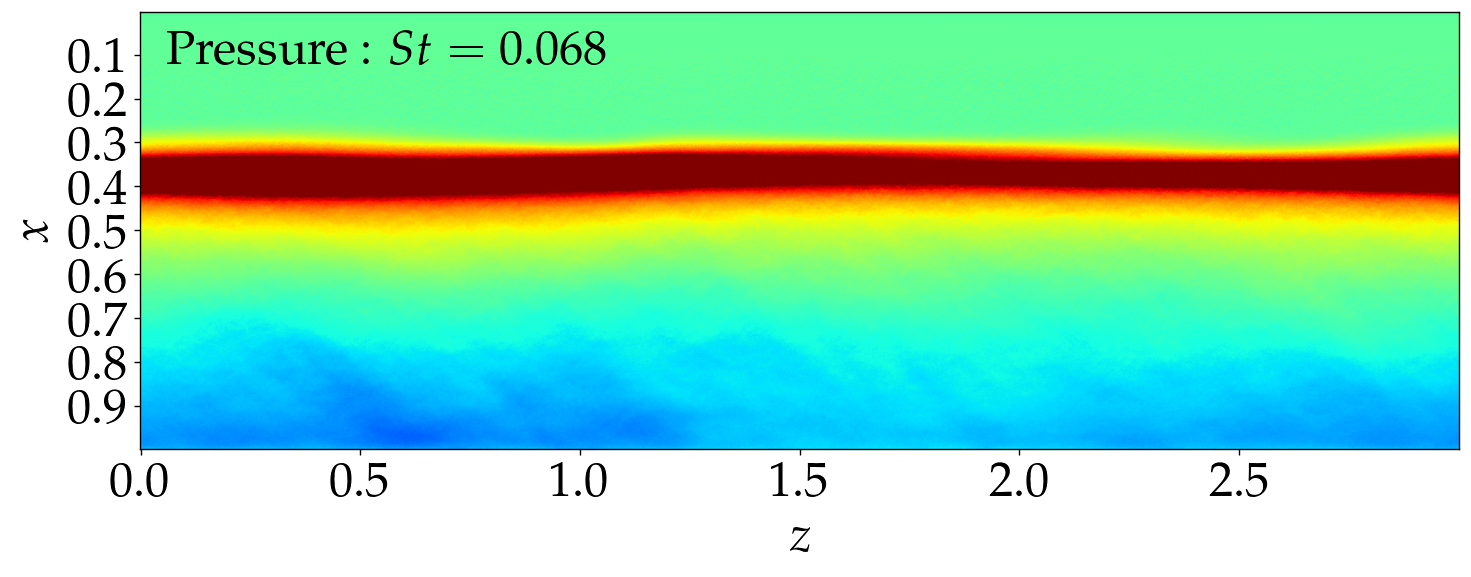}
\includegraphics[width=0.497\columnwidth]{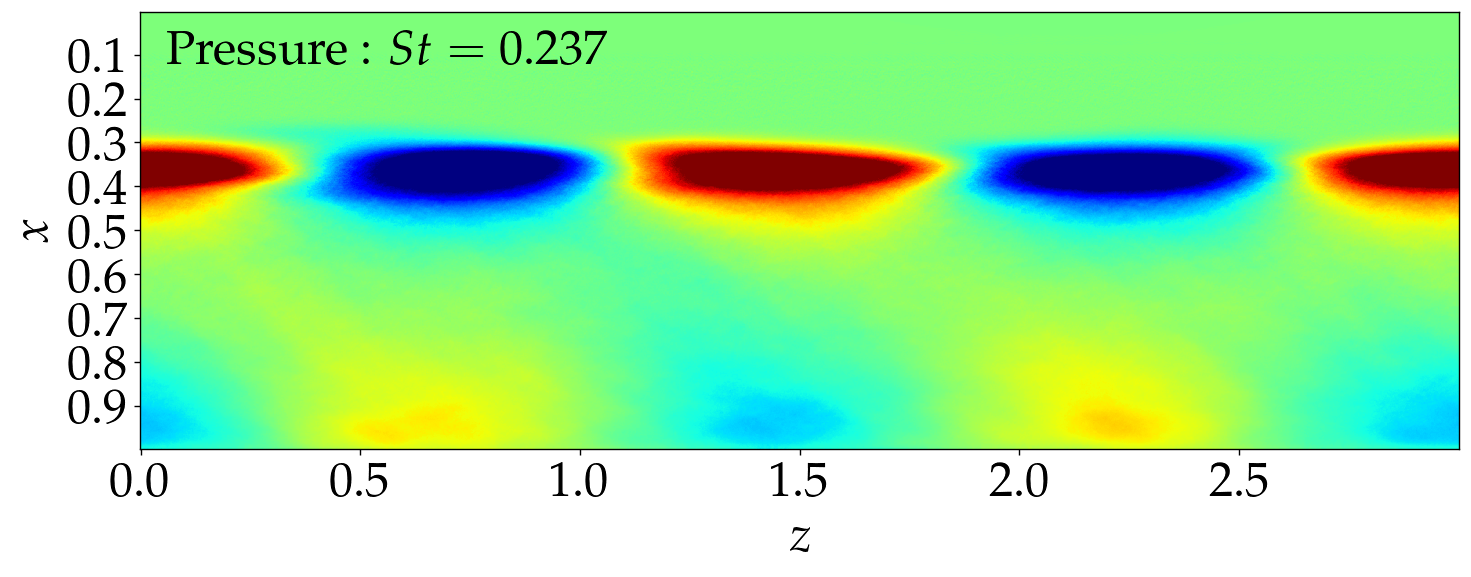}
\includegraphics[width=0.497\columnwidth]{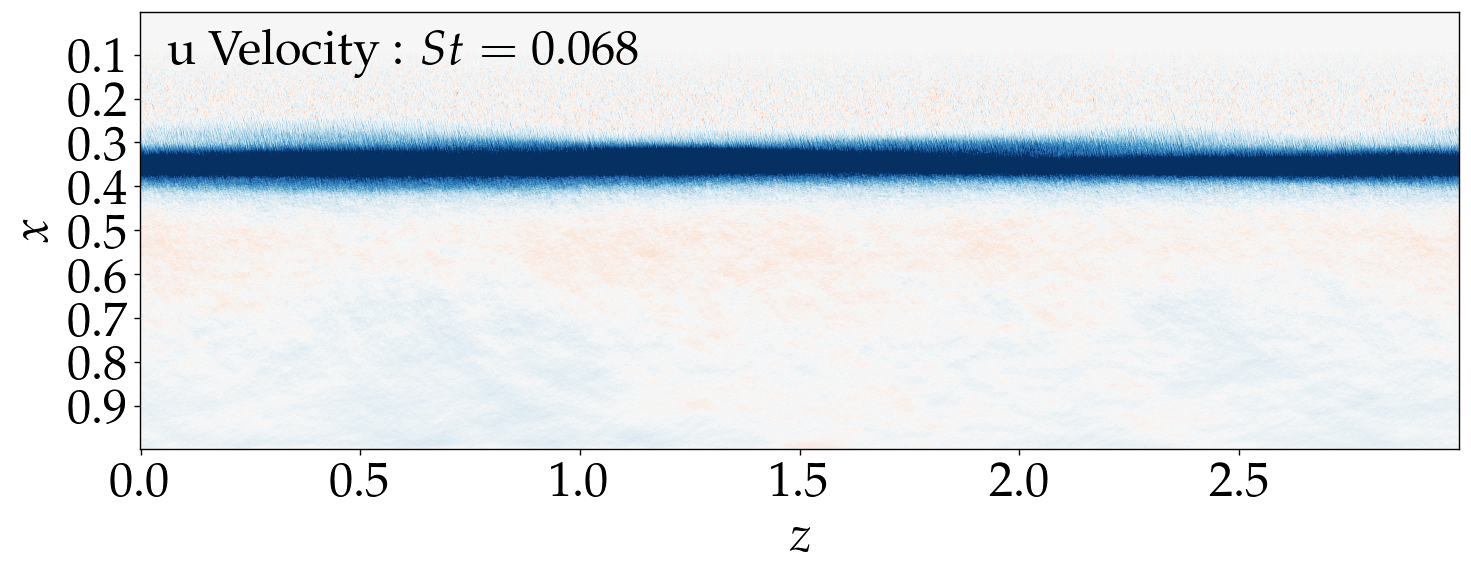}
\includegraphics[width=0.497\columnwidth]{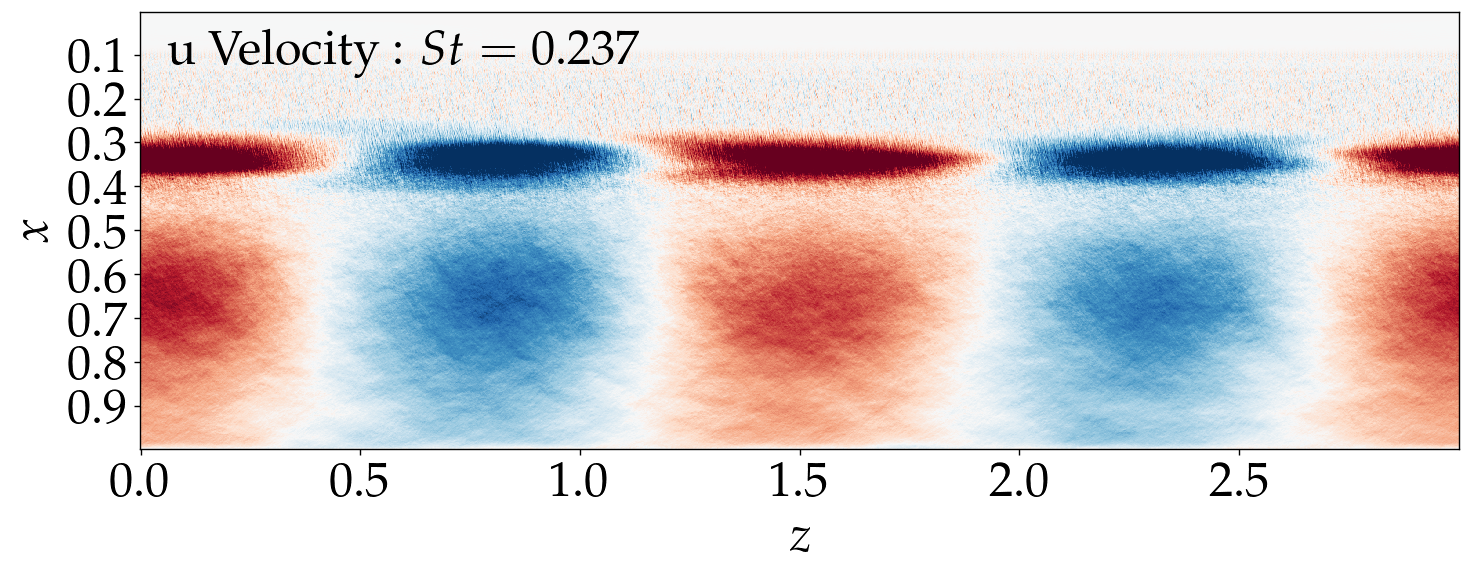}
\end{center}
\caption{SPOD modes at $\alpha = 6^{\circ}$ $AR=3$ and $\lambda = 25^{\circ}$ (as in Fig.~\ref{fig:SPOD_spectra}$(e)$ spectra), for (top) pressure and (bottom) $u$-velocity, showing the (left) low-frequency 2D shock-oscillation mode, and (right) 3D spanwise travelling buffet cell mode.}
\label{fig:SPOD_modes_AoA6_shock_and_cell}
\end{figure}

\begin{figure}
\begin{center}
\includegraphics[width=0.495\columnwidth]{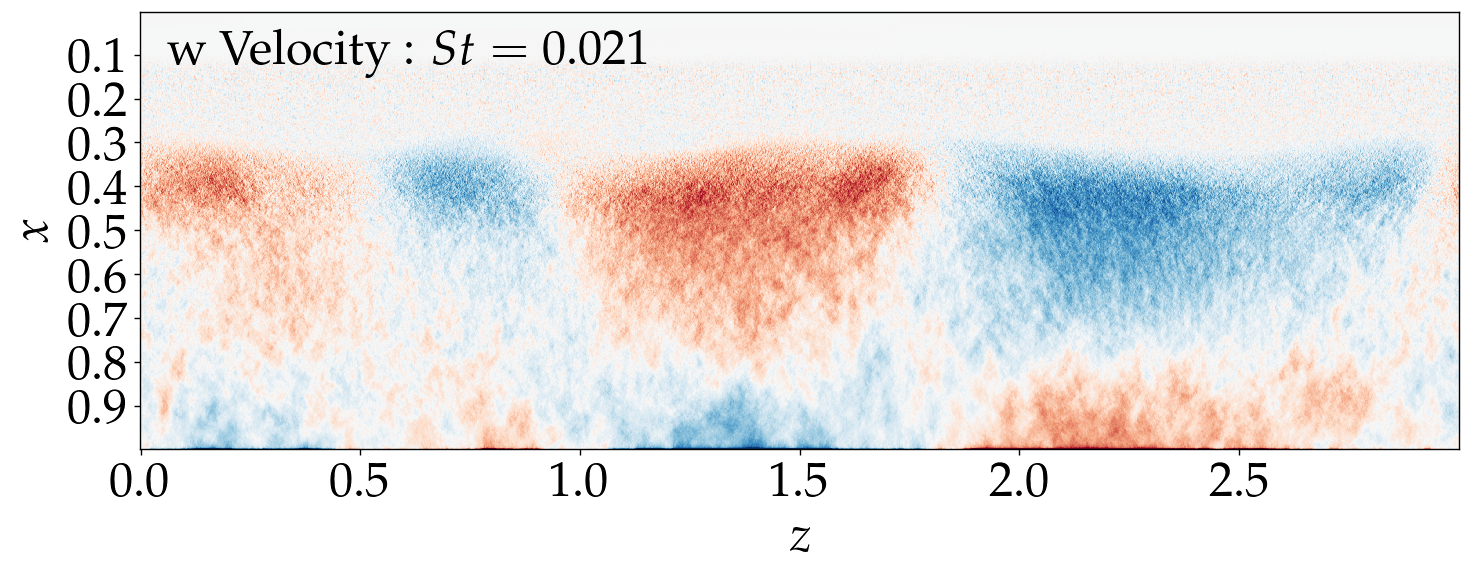}
\includegraphics[width=0.495\columnwidth]{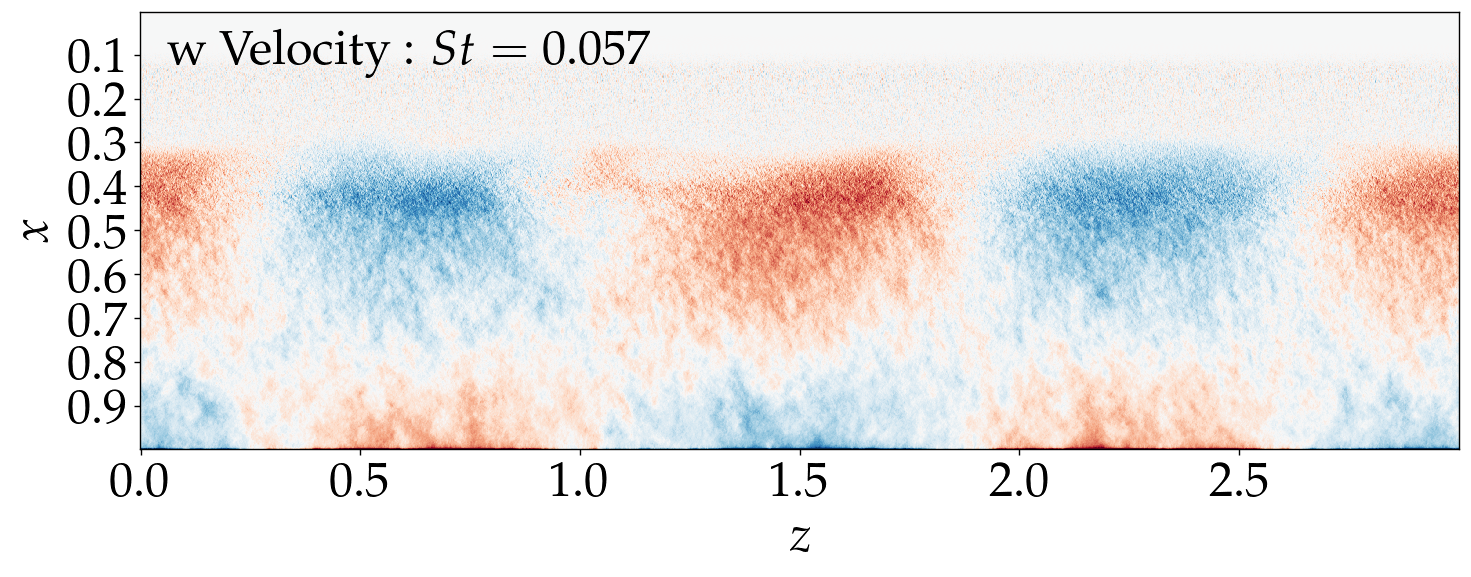}\\
\includegraphics[width=0.495\columnwidth]{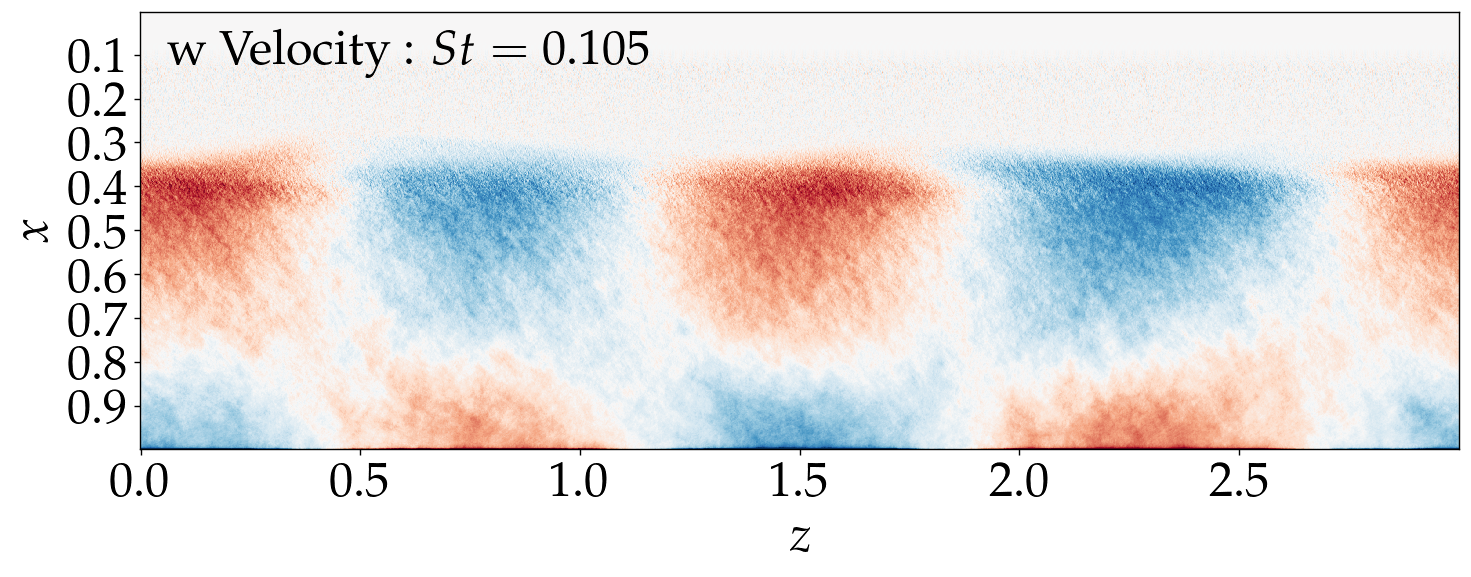}
\includegraphics[width=0.495\columnwidth]{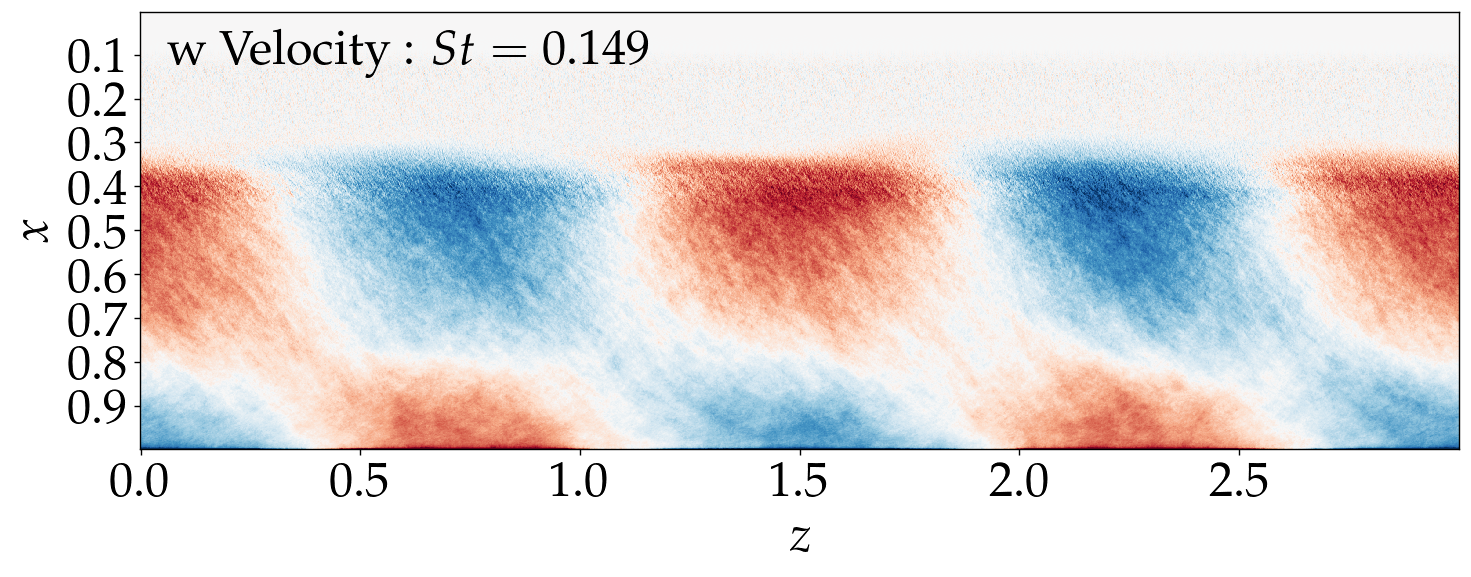}\\
\includegraphics[width=0.495\columnwidth]{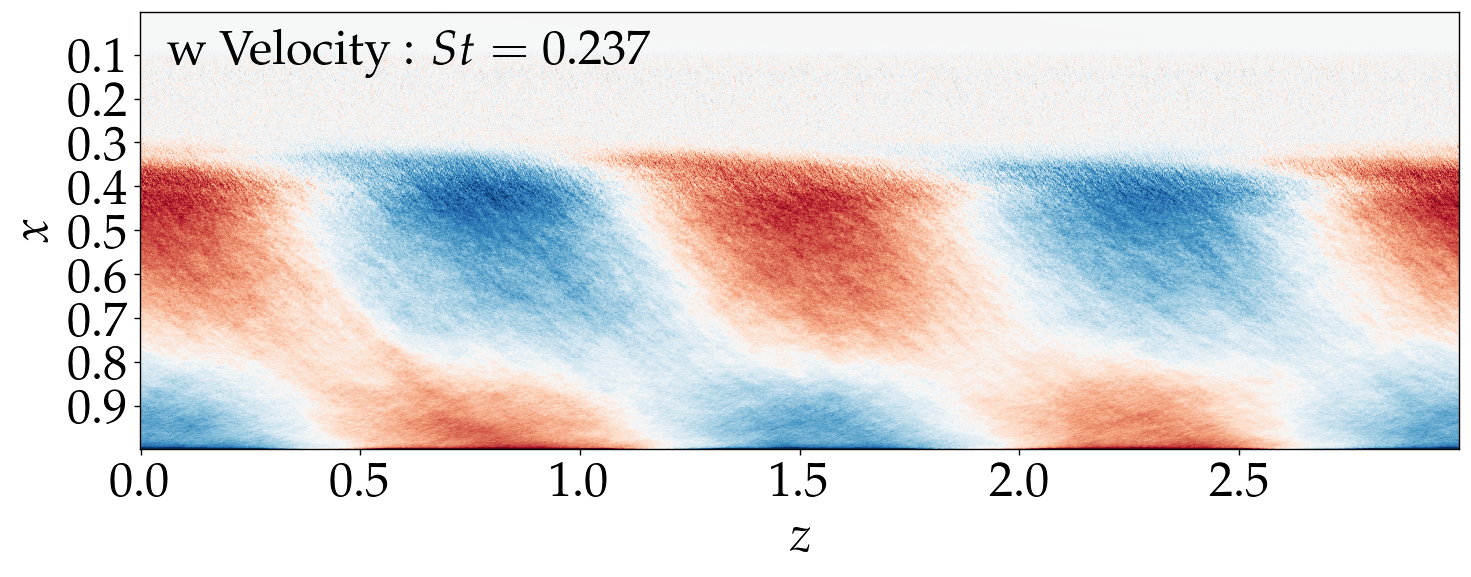}
\includegraphics[width=0.495\columnwidth]{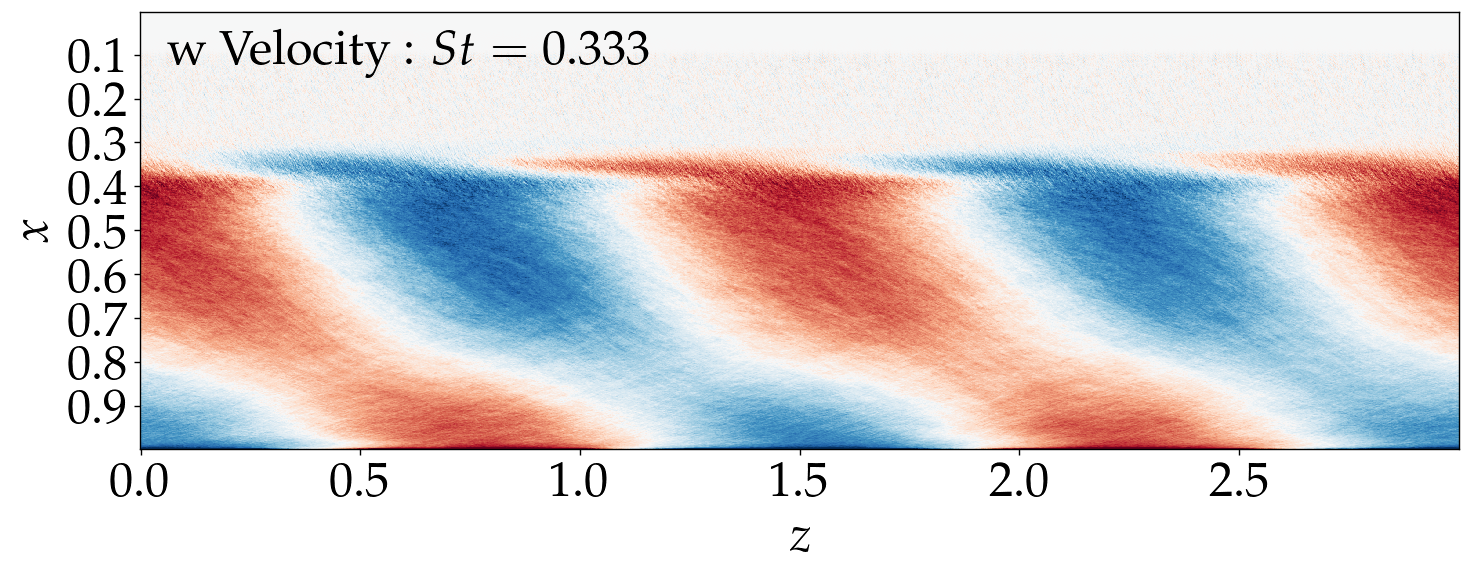}
\end{center}
\caption{SPOD modes of the $w$-velocity at $\alpha=6^{\circ}$ and $AR=3$. Showing the shift of the 3D buffet cell mode to higher Strouhal numbers for increasing sweep of (top row) $\lambda = 0^{\circ}, 5^{\circ}$ (middle row) $\lambda = 10^{\circ}, 15^{\circ}$, and (bottom row) $\lambda = 25^{\circ}, 35^{\circ}$. The unswept ($\lambda=0^{\circ}$) quasi-stationary separation cell mode \red{($St = 0.02$, \cite{lusher2025highfidelity_JFM} becomes a travelling buffet cell mode at swept conditions)}.}
\label{fig:SPOD_AoA6_w_ONLY}
\end{figure}

\red{For the pressure- and $u$-velocity-based SPOD, a low-frequency peak in the Strouhal number range of $St\approx0.08$ (indicated by the red circle) is present for all sweep angles and in agreement with the peak observed in the span-averaged $C_{L}'$-based PSD. Modes at these frequencies are usually associated to the essentially-2D shock-oscillation mode \cite{CGS2019,PBDSR2019,HT2021}. For low sweep angles (i.e. $\lambda=5,10^\circ$) and when a single peak is present, the modes around those frequencies are not strictly 2D, as shown in Fig.~\ref{fig:SPOD_modes_LAMBDA10_OVERLAP} (for brevity reported only at $\lambda=10^\circ$). As it will be shown (Fig.~\ref{fig:SPOD_AoA6_w_ONLY}), this is due to the overlap/coexistence of the 2D shock oscillation and 3D buffet cell modes in the same frequency range. This indicates that the SPOD struggles to distinguish between the two modes, suggesting that they may not be orthogonal. For medium-to-high sweep angles (i.e. $\lambda=15-35^\circ$), two distinct peaks appear in the pressure and $u$-velocity spectra. While the low-frequency spectra at 
$St\approx0.08$ is observed to be insensitive to sweep, the second peak shifts monotonically to higher frequencies up to $St=0.35$. The corresponding mode shapes at these two peaks are shown in Fig.~\ref{fig:SPOD_modes_AoA6_shock_and_cell} at $\lambda=25^\circ$ (as in Fig.~\ref{fig:buffet_cells}), clearly indicating that the low-frequency peak corresponds to the 2D span-uniform shock-oscillation mode and the high-frequency peak to the 3D buffet cell mode, in good agreement with findings in the literature \cite{Giannelis_buffet_review,OIH2018,PDL2020}.}

The 2D- and 3D-buffet modes are plotted in space in the surface contours in Fig.~\ref{fig:SPOD_modes_AoA6_shock_and_cell}. For each of these, the closest available mode from the SPOD output is selected. The 2D mode is concentrated at the shock location ($x \sim 0.375$) and is relatively uniform across the span in both the pressure and $u$-velocity decompositions, with no three-dimensionality present. In the 2D pressure mode, the shock is out of phase with the separated shear layer near the trailing edge. This is the expected 2D buffet mode found on 2D and 3D span-periodic (infinite) airfoils \cite{CGMT2009,CGS2019,LSH2024_narrow_buffet}. The 3D buffet cell mode is shown at $St=0.237$, taking the peak from the spectra in Fig.~\ref{fig:SPOD_spectra}$(e)$. A spanwise periodic mode is located at the shock with a wavelength of $\lambda_z = L_z / 2 = 3c/2$ (1.5 airfoil chords) in both flow variables. The buffet cells convect in the spanwise direction along the length of the shock (SPOD mode animations are provided in the supplementary material \cite{supplementary_material}). \red{Regarding convergence of the mode shapes with SPOD segment length, the convergence criteria \eqref{eqn:SPOD_convergence} evaluated at the target frequency of the 3D mode in Fig.~\ref{fig:SPOD_modes_AoA6_shock_and_cell} ($St = 0.237$) was found to be $0.995$ ($E_W\!=\!1.1\%$), $0.984$ ($3.2\%$), and $0.994$ ($1.2\%$) for the $(N_{seg} = 3\!\leftrightarrow\!6)$, $(N_{seg} =3\!\leftrightarrow\!13)$, and $(N_{seg} =6\!\leftrightarrow\!13)$ runs, respectively, demonstrating the excellent convergence of the mode shapes and insensitivity to the selected default segment length}.

At the trailing edge, pressure perturbations are also present with the same spanwise wavelength as the buffet cells. The trailing edge region of the 3D pressure mode is out of phase with the buffet cell along a chordwise line in $x$, and out of phase with the neighbouring cell. The convection of the buffet cell influence from shock to trailing edge in the direction of the angled cross-flow is evident. In contrast, the intensity of the $u$-velocity mode downstream of the SBLI is strongest around 65\% chord for this NASA-CRM airfoil. This point corresponds to the maximum skin-friction region downstream of the SBLI in a time/span-averaged sense (Fig.~\ref{fig:Cp_Cf}$(b)$). While the $u$-velocity mode cannot be used to comment on whether the flow is attached or separated in itself, it does highlight that SPOD performed on the streamwise velocity isolates the up/downstream moving in-phase regions of fluid along a given chordwise line. The mode in this component ignores the swept component visible in the pressure mode. Both modes traverse the spanwise direction at the same convection velocity, but demonstrate different mode shapes downstream of the SBLI depending on the choice of flow variable.

\red{For the detection and identification of sweep dependence of the buffet cell mode, SPOD analysis based on the spanwise-oriented $w$-velocity was found to be very suitable for this purpose. It can isolate the 3D mode (indicated by the green circle in Figures~\ref{fig:SPOD_spectra}$(a)$-$(f)$) and track its evolution with sweep angle, even when it overlaps in the same frequency band as the shock oscillation mode.} Figure~\ref{fig:SPOD_AoA6_w_ONLY} isolates the spanwise-oriented velocity perturbations via $w$-based SPOD at the characteristic Strouhal numbers for the buffet cell mode (Fig.~\ref{fig:SPOD_spectra}) between $\lambda = \left[0^{\circ}, 35^{\circ}\right]$. At $\lambda=0^{\circ}$, a \blue{quasi-stationary} separation cell mode is observed at $St=0.02$, as reported in \cite{lusher2025highfidelity_JFM}. Upstream travelling fluid within the separated boundary layer diverges in the $\pm z$ directions at the saddle point of the separation bubble along the shock front at $x \sim 0.35$. Two wavelengths of the instability are visible across the span. When the flow is unswept, these separation cells form at different spanwise locations between successive periods of the low-frequency buffet cycle (Figure 9, 11 \cite{lusher2025highfidelity_JFM}). The separation cells grow and reattach at a given spanwise location with minimal spanwise movement when the flow is unswept (Fig.12, \cite{lusher2025highfidelity_JFM}). When a weak cross-flow $(\lambda = 5^{\circ}, 10^{\circ})$ is imposed, the same mode becomes a spanwise travelling mode of the same wavelength but increased frequency of ($St=0.06,0.1$). At these weakly-swept conditions, the perturbations at the shock are out of phase with the trailing edge flow. While there is strong three-dimensionality from the separation cells, the 2D chordwise motion of the shock still dominates (spectra in Fig.~\ref{fig:SPOD_spectra}).

At $\lambda = 15^{\circ} - 35^{\circ}$ in the SPOD modes of Fig.~\ref{fig:SPOD_AoA6_w_ONLY}, the increased sweep is visible downstream of the SBLI. The 3D mode strengthens and shifts to the intermediate frequency range of $St \sim 0.33$. At the strongest sweep level of $\lambda = 35^{\circ}$ (as in the full aircraft swept wing configuration the 2D-CRM section in this study is taken from \cite{CRM}), the 3D mode strength is comparable to the 2D shock mode (Fig.~\ref{fig:SPOD_spectra}). This may be one explanation as to why the 3D buffet cell mode is found to be dominant on finite-wing configurations \cite{T2020,MTP2020,SH2023,VMD_Ohmichi2024}, compared to the 2D shock mode dominance on unswept/narrow straight wings.

\subsection{Further discussion and buffet cell convection models}\label{subsec:further_discussion}

\begin{figure}
\centering
\includegraphics[width=1\columnwidth]{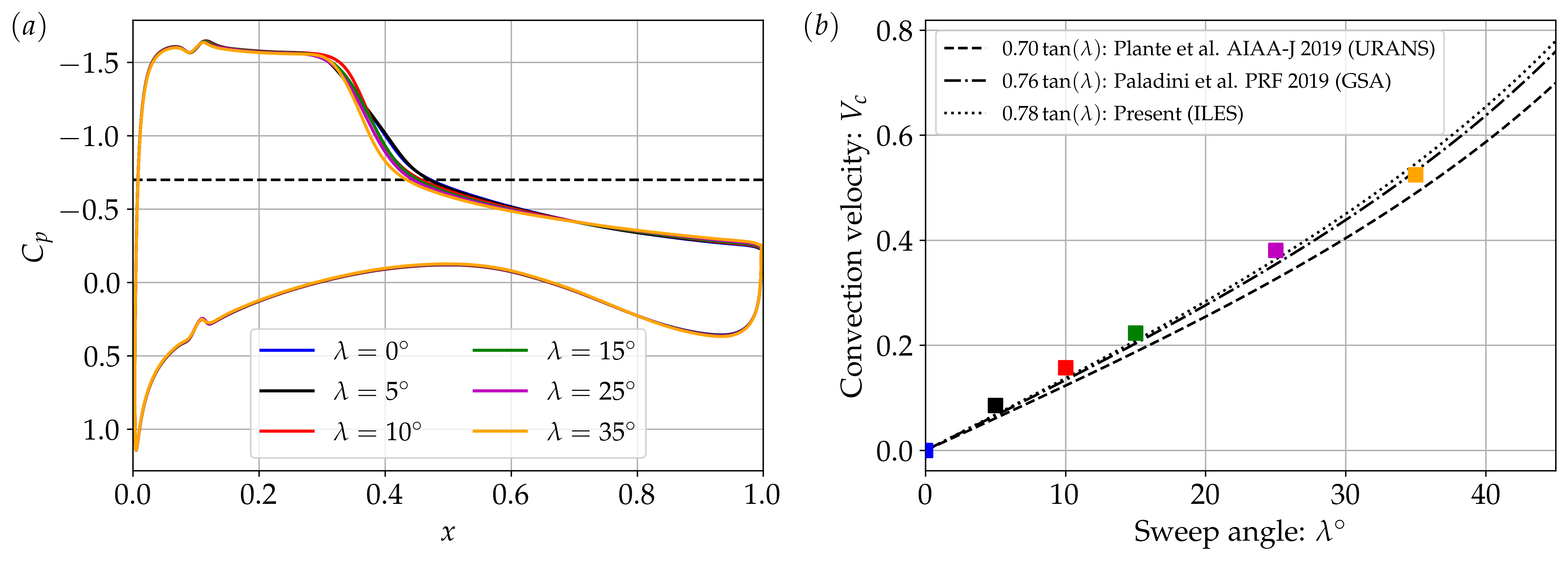}
\caption{$(a)$ span/time-averaged pressure coefficient and $(b)$ buffet cell convection velocity as functions of the sweep angle $\lambda = \left[0^{\circ}, 35^{\circ}\right]$ at $AR=3$ and $\alpha=6^{\circ}$. A best fit line and the buffet cell scaling models of Plante et al \cite{PDL2020,PDBLS2021} and Paladini et al. \cite{PBDSR2019} are compared.}
\label{fig:convection_velocity}
\end{figure}

Figure~\ref{fig:convection_velocity}$(a)$ shows the span- and time-averaged pressure coefficient over the full range of sweep angles. The pressure side of the airfoil is found to be entirely insensitive to sweep effects for this buffet configuration in the mean. On the suction side, the transition region ($x \sim 0.1$) and region adjacent to the trailing edge ($x \sim 0.9$) also overlap in the mean. The region occupied by the buffet cells $0.3 < x < 0.5$ shows the strongest sensitivity to sweep, where the 2D chordwise shock oscillations \red{coexist} with the unsteady spanwise propagation of the separation cells (Fig.~\ref{fig:buffet_cells}). This result is consistent with the swept airfoil URANS cases of Plante et al \cite{PDL2020,PDBLS2021}, who showed the same $C_p$ sensitivity to sweep angle was localised only to the region around the shock.

Regarding the open question of the importance of shock waves in buffet, other authors \citep{PDL2020,PDBLS2021} have previously highlighted similarities between three-dimensionality on airfoils in the form of buffet cells (observed under transonic flow conditions with shock waves and moderate-AoA) and those of stall cells (low subsonic flow and high-AoA). Both cases correspond to situations where the flow is highly-separated at a given time instance, and, for sufficiently wide aspect wings ($AR \gtrsim 1$ \citep{lusher2025highfidelity_JFM}), 3D cellular separation patterns naturally develop across the span width. The buffet cells in the present study (Fig.~\ref{fig:buffet_cells}) are coherent regions of flow reversal that are concentrated at the shock, but also extend downstream to the trailing edge. In-between the cells in the spanwise direction are regions of attached flow. The time/span-averaged distributions of $C_p$ in Fig.~\ref{fig:convection_velocity}$(a)$ show that the separated trailing edge region is essentially independent of sweep when span-averaging is applied. All of the wall pressure variations that persist (due to not having an infinitely long time series) are localised to the shocked region of the flow ($0.3 < x < 0.5$), where the 3D separation cells \red{coexist} with the chordwise motion of the shock wave. Here, the greatest variance is observed for differing levels of crossflow. Although similarities do undoubtedly exist between high-speed buffet cells and low-speed stall cells \citep{PDBLS2021}, at transonic speeds, the greatest sensitivity to cross-flow shows up primarily at the shock (see, also the clear concentration of pressure fluctuation energy at the shock in the SPOD modes in Fig.~\ref{fig:SPOD_modes_AoA6_shock_and_cell}).

\red{The convection velocity of the perturbations is a commonly reported metric \citep{PBDSR2019, PDL2020} used to compare buffet cell characteristics between both simulation and experiment, as a way of assessing generality and agreement between different flow conditions and airfoil and wing geometries}. Here, the characteristic frequency of the 3D modes from the SPOD are used to calculate the spanwise convection velocity of the buffet cell disturbances along the shock front as a function of the sweep angle. The convection velocity $V_c$ is related to the Strouhal number $St$ as
\begin{equation}\label{eqn:convection}
V_c = \lambda_z \cdot f = \lambda_z \frac{St \cdot U_{\infty,2D}}{c},
\end{equation}
based on the airfoil chord of $c=1$. The wavelength of the perturbation, $\lambda_z$, is found to be independent of the sweep angle, $\lambda$, and is set by the aspect ratio ($AR = L_z / c = 3$) of the spanwise-periodic airfoil as $\lambda_z = L_z / 2 = 1.5 c$ at the present flow conditions and geometry. The observed independence of the 3D mode wavelength with sweep is consistent with the tri-global stability-based study of buffet on infinite wings by \citet{HT2021}, who found the mode to have a wavelength equal to one airfoil chord ($\beta\approx 2\pi$) in their cases. Similarly, Crouch et al. \cite{CGS2019} and Paladini et al. \cite{PBDSR2019} also determined the GSA dominant modes to have wavelength corresponding to one chord. It should be noted that for experiments on full-aircraft configurations, Masini et al. \cite{MTP2020} and Sugioka et al. \cite{SNKNNA2021} determined the wavelenghth of the outboard travelling buffet cells to be 0.6-1.2 and 1.3 mean aerodynamic chords, respectively. Noting the similarities between the dominant mechanisms seen on span-periodic and finite wings, this agreement provides additional confidence to the present results.

In agreement with our results and those of \citet{PDBLS2021}, \citet{T2020} also reported an increased frequency of the mode as a function of the applied sweep angle. The convection velocities from our results are plotted as a function of sweep angle in Fig.~\ref{fig:convection_velocity}$(b)$. Comparisons are made to the URANS results on the OALT25 airfoil at a chord-based Reynolds number of $Re_c = 3\times10^6$ and aspect ratio of $AR=6$ by Plante et al \cite{PDL2020,PDBLS2021}, who proposed a buffet cell model defined as
\begin{equation}\label{eqn:plante}
V_c = 0.7 \tan\left(\lambda\right).
\end{equation}
and the GSA-based evaluation of the convective velocity by Paladini et al. \cite{PBDSR2019}, which is calculated using the wavenumber of the dominant 3D buffet mode in their case.

Despite the differences in airfoil geometry (NASA-CRM vs OALT25), Reynolds number ($Re_c = 0.5 \times10^6$ vs $Re_c = 3.0 \times10^6$), aspect ratio ($AR=3$ vs $AR=6$), and simulation fidelity (ILES vs URANS), the present data matches the proposed model well. A similar trend between buffet cell convection velocity and sweep angle is observed. The model slightly underestimates the convection velocities for the present cases, most likely due to the above differences in configuration and flow conditions. For the present data, a best fit relationship is found to be $V_c = 0.78\tan\left(\lambda\right)$. This amounts to only a $\sim 10\%$ difference in the scaling coefficient despite the multiple differences between the two simulation campaigns. The work of Paladini et al. \cite{PBDSR2019} found a convection velocity relationship of $V_c = 0.76\tan\left(\lambda\right)$ in global stability-based computations of the OAT15A airfoil at $Re_c = 3.2\times 10^6$, even closer to our result.

It should be noted that for span-periodic wings, the spanwise domain width imposes discrete admissible wavenumbers (i.e. submultiples of $L_z$) for the 3D instability. Although determining the dominant wavenumber is out of scope of our scale-resolving method that would require a large number of simulations with different domain spanwise widths, the agreement with Plante et al \cite{PDL2020,PDBLS2021} and Paladini et al. \cite{PBDSR2019} - for which the dominant wavenumbers were used to calculate the convective speeds - suggests that the present domain is large enough to accommodate and provide insight into the characterization of the 3D buffet cell mechanism.

In summary, these results demonstrate that the 3D component of buffet is largely a separation bubble-based instability \red{that coexists with the} chordwise shock motion under certain flow conditions. At zero to low sweep strengths ($\lambda \leq 5^{\circ}$), the 3D mode appears at frequencies below the 2D shock mode frequency, and the chordwise 2D oscillations are found to be dominant. At higher sweep angles, \red{the 3D mode shifts to higher frequencies and is concentrated primarily to the shocked region} (SPOD mode in Fig.~\ref{fig:SPOD_modes_AoA6_shock_and_cell}), with the relative energy content of the 3D separation mode becoming comparable to the 2D shock mode (SPOD spectra in Fig.~\ref{fig:SPOD_spectra} and $C_{L}^{\prime}$ PSD in Fig.~\ref{fig:deg6_lines_low}$(d)$). At sweep angles commonly used in the design of wings of modern commercial aircraft ($\lambda \sim 35^{\circ}$), the 3D mode is dominant and localised along the shock front. Given that the 3D buffet mode becomes dominant at high sweep angles even on idealised periodic straight wings, it is understandable that the 2D shock mode is difficult to observe on finite wings with the additional inherent geometrical complexity imposed on the problem. \red{As an extension of this research, the relation between these two instabilities should be interrogated using advanced analysis techniques such as resolvent analysis \cite{ModalDecomp_Taira2017}, bi-spectral modal analysis \citep{bmd2020} or information-theoretic causal inference \cite{HATAYAMA2026}.}

\section{Conclusions}\label{sec:conclusions}

A set of ten large-scale high-fidelity ILES simulations of considerable computational cost (on $N \sim 8 \times 10^{9}$ mesh points) has been performed to investigate the effect of sweep angle on both 2D and 3D aspects of the turbulent transonic shock buffet phenomenon. Flow databases were generated to enable data-driven modal analysis of the 3D buffet phenomena. \blue{In addition to explicitly resolving the turbulence to capture the effects of small-scale fluctuations on the large-scale phenomena, a key objective of the study was to investigate the role of boundary‑layer separation on the emergence of spanwise‑varying (3D) buffet cells. URANS/RANS predictions are known to be sensitive to turbulence‑model form and numerical settings in separated, unsteady shock boundary layer interactions}.

Cases corresponding to minimally separated ($\alpha=5^{\circ}$), and largely separated ($\alpha=6^{\circ}$) flow (in a time/span-averaged sense at the shock) were compared, with the former proving to exhibit essentially-2D span-uniform dynamics at the shock front, without the presence of 3D `buffet cells' \citep{IR2015,Giannelis_buffet_review}. Patches of intermittent separation were, however, observed adjacent to the trailing edge during periods of maximum flow separation (Fig.~\ref{fig:AoA5_surface_contours}), consistent with previous studies which argued that high-speed buffet-cells and low-speed stall-cells \citep{PDL2020,PDBLS2021}, may share a common physical origin. The separation cells in this case did not strongly interact with the SBLI nor cause \blue{macroscopic} protrusions along the shock front.

As the mean flow separation increases at a higher angle of incidence ($\alpha=6^{\circ}$), clear 3D buffet cells are observed with a spanwise wavelength of $\lambda_z=1-1.5c$ airfoil chords. Furthermore, the \blue{intermittent quasi-stationary} 3D separation cell mode identified on unswept infinite wings in \citet{lusher2025highfidelity_JFM} (at $St = 0.02$), was shown to become a spanwise travelling mode and shift to intermediate frequencies ($St = \left[0.06, 0.35\right]$) as sweep is imposed. The disturbances convect in the spanwise direction along the shock front, consistent with the global stability analysis conclusions of \citet{CGS2019}. \orange{Due to the wing sweep angle, the 3D mode is oblique with velocity components in both the streamwise and spanwise directions}. This set of simulations importantly links the \blue{intermittent quasi-stationary} mode on unswept wings to the more commonly reported travelling buffet cell mode for the first time via high-fidelity simulation and modal analysis, independently from RANS-based simulation and stability calculations \blue{which are often inaccurate in highly-separated boundary layers}. 

An SPOD modal analysis was performed on the obtained datasets. \blue{The spectral analysis showed that the unsteady aerodynamic loads are due to a superposition of the distinct 2D chordwise low-frequency shock oscillations and spanwise convection of the separation-based buffet/stall cells}. Characteristic frequencies were isolated from each case. By performing modal analysis on different flow quantities, the SPOD modes based on the $w$-velocity component proved essential to clearly distinguish the energy peaks related to the intermediate-frequency ($St=0.06-0.35$) cellular 3D buffet cell mode from those of the nominally-2D low-frequency ($St=0.08$) shock-oscillation mode, even in cases where they overlap (e.g. $\lambda = 10^{\circ}$). The 2D mode was shown to be insensitive to sweep, whereas the frequency associated with 3D buffet cells was monotonically proportional to the magnitude of the imposed sweep angle. In good agreement with previous studies \citep{HT2021}, the wavelength of the 3D mode was shown to be independent of sweep, being instead set by the aspect ratio of the spanwise periodic computational domain. \orange{At a sweep of $\lambda=25^{\circ}$ the shock and buffet cell modes were found to have similar energy levels, in general agreement with the higher Reynolds number experiments of \citet{DSO2022}. Furthermore, the $St=0.105$ frequency of the 3D mode at $\lambda=10^{\circ}$ agreed very well with the value obtained from experiments at the same sweep angle by \citet{SKK2022} on the same NASA-CRM airfoil profile. This was despite the higher Reynolds number of $Re_c=2.5\times 10^{6}$ in their case, and the three-dimensionality imposed by the wind tunnel walls}.

The pressure-based spectral energy level of the 3D mode increases with sweep and is larger than the 2D shock mode for sweep angles comparable to those seen on wings in full-aircraft configurations (corresponding to $\lambda=35^{\circ}$). In addition to finite-wing effects (as shown by \cite{DSO2021,DSO2022}), these can serve as an explanation for the apparent absence of 2D shock modes in fully 3D wings \cite{T2020,MTP2020,SH2023,VMD_Ohmichi2024}. To further strengthen the similarities between transonic buffet and low-speed stall, the convection velocity of the buffet cells agrees well with scaling law models proposed by \citet{PDL2020,PDBLS2021}. The three-dimensionality of buffet is found to be primarily a separation based mechanism \red{that coexists with} the chordwise shock wave motion under certain conditions, with mean flow separation around the shock determined to be a necessary condition for the 3D mode to become dominant. \red{Future work will focus on the application of more advanced analysis techniques to the database to investigate the interplay between the modes}, and the inclusion of no-slip sidewall effects to better understand the flow confinement interaction between corner/main separations found in wind tunnel experiments and at the wing/fuselage junction in aircraft.

\begin{acknowledgments}
D.J.L was supported by a JSPS KAKENHI grant (22F22059). Computational time was provided by the JSS3 supercomputer at JAXA Chofu Aerospace Center, and the Fugaku supercomputer on HPCI project hp220226. We thank Dr. Yuya Ohmichi (JAXA) for discussions on this work.
\end{acknowledgments}



\bibliography{apssamp}

\end{document}